\documentclass{aa}
\usepackage{txfonts}

\usepackage{natbib,twoopt}
\usepackage[breaklinks=true]{hyperref} %% to avoid \citeads line fills
\bibpunct{(}{)}{;}{a}{}{,}             %% natbib format for A&A and ApJ
\makeatletter
 \newcommandtwoopt{\citeads}[3][][]{\href{http://adsabs.harvard.edu/abs/#3}%
   {\def\hyper@linkstart##1##2{}%
    \let\hyper@linkend\@empty\citealp[#1][#2]{#3}}}
 \newcommandtwoopt{\citepads}[3][][]{\href{http://adsabs.harvard.edu/abs/#3}%
   {\def\hyper@linkstart##1##2{}%
    \let\hyper@linkend\@empty\citep[#1][#2]{#3}}}
 \newcommandtwoopt{\citetads}[3][][]{\href{http://adsabs.harvard.edu/abs/#3}%
   {\def\hyper@linkstart##1##2{}%
    \let\hyper@linkend\@empty\citet[#1][#2]{#3}}}
 \newcommandtwoopt{\citeyearads}[3][][]%
   {\href{http://adsabs.harvard.edu/abs/#3}
   {\def\hyper@linkstart##1##2{}%
    \let\hyper@linkend\@empty\citeyear[#1][#2]{#3}}}
\makeatother
\usepackage{graphicx}   % Including figure files
\usepackage{subfigure}

\usepackage{multirow, makecell}
\usepackage{booktabs}

\usepackage[normalem]{ulem} 
\DeclareUnicodeCharacter{2212}{-}

\usepackage{placeins}

\def\msun{$h^{-1}$~\mbox{M$_{\sun}$}} %Solar mass

\begin{document}

% Title of the paper, and the short title which is used in the headers.
% Keep the title short and informative.
%\title[Host halo mass growth on low-mass central galaxies in the vicinity of massive systems]{Host halo mass growth on low-mass central galaxies in the vicinity of massive systems}
\title{The evolution of low-mass central galaxies in the vicinity of massive structures and its impact on the two-halo conformity}

   %\subtitle{I. Overviewing the $\kappa$-mechanism}

   \author{Daniela Palma
          \inst{1} \and
          Ivan Lacerna \inst{1,2}%\fnmsep\thanks{Just to show the usage
          %of the elements in the author field}
          \and
          M. Celeste Artale
          \inst{3}
          \and
          Antonio D. Montero-Dorta,
          \inst{4}
          \and
          Andrés N. Ruiz
          \inst{5,6}
          \and
          Sofía A. Cora
          \inst{7,8}
          \and
          Facundo Rodriguez
          \inst{5,6}
          \and
          Diego Pallero
          \inst{9}
          \and
          Ana O'Mill
          \inst{5,6}
          \and
          Nelvy Choque-Challapa
          \inst{1,4}
          }

   \institute{Instituto de Astronom\'ia y Ciencias Planetarias de Atacama, Universidad de Atacama, Copayapu 485, Copiap\'o, Chile.\\
              \email{daniela.palma@postgrados.uda.cl}
         \and
             Millennium Institute of Astrophysics, Nuncio Monsenor Sotero Sanz 100, Of. 104, Providencia, Santiago, Chile. 
         \and
             Universidad Andres Bello, Facultad de Ciencias Exactas, Departamento de Ciencias Físicas, Instituto de Astrofísica, Av. Fernández Concha 700, Santiago, Chile.
        \and
            Departamento de Física, Universidad Técnica Federico Santa Maria, Av. Vicuña Mackenna 3939, 8940897, San Joaquín, Santiago, Chile.
        \and
            CONICET. Instituto de Astronomía Teórica y Experimental (IATE). Laprida 854, Córdoba X5000BGR, Argentina.
        \and 
            Universidad Nacional de Córdoba (UNC). Observatorio Astronómico de Córdoba (OAC). Laprida 854, Córdoba X5000BGR, Argentina.  
        \and 
            Instituto de Astrof\'isica de La Plata (CCT La Plata, CONICET, UNLP), Paseo del Bosque s/n, La Plata, Argentina.
        \and 
            Facultad de Ciencias Astron\'omicas y Geof\'isicas, Universidad Nacional de La Plata, Paseo del Bosque s/n, La Plata, Argentina.
        \and 
            Departamento de F\'isica, Universidad T\'ecnica Federico Santa Mar\'ia, Avenida Espa\~na 1600, Valpara\'iso, Chile.
            }

   \date{Received --; accepted --}

% The list of authors, and the short list which is used in the headers.
% If you need two or more lines of authors, add an extra line using \newauthor

% Enter the current year, for the copyright statements etc.
%\pubyear{2024}

\abstract
{
We investigated the population of low-mass central galaxies with M$_{\star}= 10^{9.5} - 10^{10}$ $h^{-1}$ M$_{\odot}$, inhabiting regions near massive groups and clusters of galaxies using the IllustrisTNG300 and MDPL2-SAG simulations. We set out to study their evolutionary histories, aiming to find hints about the large-scale conformity signal they produce. 
We also used a control sample of central galaxies with the same stellar mass range located far away from massive structures. For both samples, we find a subpopulation of galaxies accreted by another halo in the past, but now considered central galaxies; we refer to these objects as former satellites. The number of former satellites is higher for quenched central galaxies near massive systems, with fractions of 45\% and 17\% in IllustrisTNG300 and MDPL2-SAG, respectively. 
The differences in the numerical resolution of each simulation lead to the different fractions of former satellites.
Our results in TNG300 show that former satellites ``pollute'' the sample of central galaxies because they suffered environmental processes when they were satellites hosted typically by massive dark matter halos (M$_{200} \geq$ $10^{13}$ $h^{-1}$ M$_{\odot}$) since $z \lesssim 0.5$.
After removing former satellites, the evolutionary trends for quenched central galaxies near massive structures are fairly similar to those of the quenched control galaxies, showing small differences at low redshift. For MDPL2-SAG instead, former satellites were hosted by less massive halos, with a mean halo mass around $10^{11.4}$ $h^{-1}$ M$_{\odot}$, and the evolutionary trends remain equal before and after removing former satellite galaxies. We also measured the two-halo conformity, that is, the correlation in the specific star formation rate between low-mass central galaxies and their neighbors at megaparsec scales, and how former satellites contribute to the signal at three different redshifts: $z=0, 0.3,$ and 1. The time evolution of the conformity signal in the simulations presents apparent contradictory results: it decreases from $z=0$ to $z=1$ in MDPL2-SAG, while it increases in TNG300. 
However, after removing former satellites in the latter, the signal is strongly reduced, but practically does not change at $z \leq 0.3$, and it disappears at $z=1$. 
We compare our findings with recent literature data and discuss the conformity measurements, as different approaches can lead to varying results.
}


\titlerunning{On the evolution of low-mass central galaxies}
\authorrunning{Daniela Palma et al.}
\maketitle

%%%%%%%%%%%%%%%%%%%%%%%%%%%%%%%%%%%%%%%%%%%%%%%%%%

%%%%%%%%%%%%%%%%% BODY OF PAPER %%%%%%%%%%%%%%%%%%

\section{Introduction} \label{sec:intro}
The processes that cause the quenching of star formation in galaxies are still not fully understood.
On the one hand,  there are internal effects related to the feedback produced by supernovae (SNe) and active galactic nuclei (AGNs)  that contribute to the star formation quenching. On the other hand, we know that several environmental effects can occur inside galaxy clusters, such as ram pressure stripping (RPS, \citealt{GunnandGott_1972}; \citealt{FormanandJones1982}; \citealt{GiovanelliandHaynes_1985}; \citealt{Aragonsalamanca_1993}), which can strip the hot and cold gas, being stronger at distances closer to the halo center. Galaxy encounters or harassment (\citealt{Moore_1996}; \citealt{Moore_1998}) are also likely to shut down the star formation activity. Starvation,  when the accretion of gas is ceased due to a lack of gas supply (\citealt{Larson_1980}; \citealt{McCarthy_2008}; \citealt{Peng_2015}), and tidal forces produced by the potential well of the halo (\citealt{Moore_1999}; \citealt{Newberg_2002}; \citealt{McCarthy_2008}; \citealt{Villalobos_2014}) are other environmental effects that can also contribute. 
\cite{ManBelli_2018} provide a concise overview and a comprehensive classification of the quenching mechanisms in massive galaxies, highlighting the diverse physical processes that disturb the balance between gas accretion and ejection, consequently impacting the reservoir of cold gas available for star formation.

Some environmental effects can extend to the vicinity of clusters up to scales of several megaparsecs; an example of this is pre-processing (\citealt{Fujita_2004, Wetzel_2013, Haines_2015, Pallero_2019, Hough_2023, Piraino-Cerda_2024}) that occurs before galaxies enter galaxy clusters, showing a reduction in their star formation activity. The main mechanism responsible for pre-processing is RPS, which acts within clusters or groups (\citealt{Tecce_2010}; \citealt{Pallero_2022}), and also outside galaxy clusters (e.g., \citealt{Bahe_2013}; \citealt{Ayromlou_2021}). 
Galaxies may undergo pre-processing once or several times during their lifetimes, leading to the removal of hot (\citealt{Balogh_2000}) and cold (\citealt{Abadi_1999}) gas reservoirs. The efficiency of this removal can be associated with the mass of the halo that hosts these galaxies \citep{Paulo_Lopez_2024}. 
\cite{Bahe_2013} studied the environmental effects outside galaxy clusters in the \texttt{\small{Galaxies-Intergalactic Medium Interaction Calculations}} (GIMIC, \citealt{Crain_2009})
cosmological hydrodynamic simulations at $z \sim 0$. In particular, they considered galaxies with stellar masses within the stellar mass range log$_{10}$(M$_{\star}$/M$_{\odot}$) = [9.0, 11.0]. 
They found a fraction of pre-processed galaxies located at large cluster-centric distances (up to four virial radii) for the studied mass range. In addition, they showed that RPS is efficient in removing the hot gas halo around all the galaxies studied up to several virial radii (see also \citealt{Zinger_2018} for a discussion of the quenching of satellites at the outskirts of clusters).  In contrast, only for less massive galaxies (log$_{10}$(M$_{\star}$/M$_{\odot}$) $<$ 10$^{9.5}$), the RPS can also remove the cold gas near the massive clusters.
\cite{Paulo_Lopez_2024} find that galaxy groups near clusters in the Sloan Digital Sky Survey (SDSS) can quench star formation activity in galaxies even at cluster-centric distances of up to five virial radii. Their findings demonstrate a dependence of the pre-processing effect on group mass, with massive halos leading to a more significant suppression of star formation activity compared to low-mass halos (see also \citealt{Wang_Mo_Jing_2007}; \citealt{Hahn_2009}). In this scenario, less massive galaxies are more susceptible to suffering these large-scale environmental effects before entering a more massive group or cluster (\citealt{Bahe_2013, Ayromlou_2021}). 
\cite{Salerno_2022} studied different galaxy environments to analyze the star formation quenching using the MDPL2-SAG galaxy catalog \citep{Knebe_2018}, obtained from the combination of the semi-analytic model {\sc sag} \citep{Cora_2018} and the dark matter only cosmological simulation MULTIDARK P{\sc lanck} 2 \citep[MDPL2,][]{Klypin_2016}. 
In particular, they considered three galaxy types related to their location: cluster galaxies (CGs), galaxies in the filamentary infall region (FRG), and in the isotropic infall region (IRG) at three different redshifts: $z=0, 0.65$, and $0.85$. They found that the red fraction predominates for FRG over IRG at cluster-centric distances greater than $\sim 3$ virial radii, regardless of stellar mass. Furthermore, they observed that star formation quenching is more pronounced in filamentary regions than in isotropic infall regions across the entire redshift range analyzed.
Additionally, they analyzed the evolution of hot gas for these galaxies, finding that hot gas in infall and filamentary regions increases by up to $z \sim 0.5$ and then decreases. In contrast, field galaxies show a monotonic increase with a slight decrease at $z=0$. Their results showed the importance of filamentary regions in acting more efficiently in pre-processing galaxies along their evolution (see also \citealt{Sarron_2019}).

In the literature, when a galaxy crosses the virial radius of a more massive system (e.g., a galaxy cluster) once or several times during its lifetime, and it is later located outside the virial radius of that dark matter halo, it is  commonly referred to as a backsplash, fly-by, or former satellite galaxy (e.g., \citealt{Salerno_2022}; \citealt{Ayromlou_2023} \citealt{Hough_2023}; \citealt{Ruiz_2023}). The first two definitions consider the trajectory of galaxies that cross the virial radius (see \citealt{Ayromlou_2023}) and at some future time fall back into the potential well of the cluster are backsplash galaxies; fly-by galaxies, on the other hand, are galaxies that, due to their elliptical orbits, passed the virial radius only once. We widely refer to the galaxies in both cases as former satellites.

\cite{Hough_2023} used the MDPL2-SAG galaxy catalog to study four populations of galaxies based on their orbits in and around relaxed clusters: recent infallers, ancient infallers, backsplash galaxies, and neighbor galaxies. They found that around $\sim$65\% of quenched galaxies are backsplash galaxies. These galaxies constitute the dominant population in the outskirts of clusters (up to $\sim 3$ virial radii).  
Their findings suggest that around 70\% of passive backsplash galaxies experienced pre-processing effects. On the contrary, $\sim$65\% of star-forming backsplash galaxies did not show evidence of pre-processing effects. In \cite{Ruiz_2023}, the quenching of star formation is studied at different stages of the dynamical evolution of a galaxy. They found that the environmental effects of a cluster can affect galaxies that cross the virial radius of the cluster after a single passage, in which low-mass galaxies ($<$ 3$\times 10^{10}$ $h^{-1}$ M$_{\odot}$ at $z=0$) orbiting close to the cluster center are the most affected.

Most of the above-cited works  on galaxies in the outskirts of clusters do not distinguish between central and satellite galaxies, where the centrals reside near the center of the potential well of host dark matter halos. Central galaxies are hosted by halos of different masses, including low-mass systems.
\cite{Wetzel_2012} tested how the fraction of quenched galaxies from the SDSS Data Release 7 (DR7, \citealt{Blanton_2005}; \citealt{Abazajian_2009}) changes at different cluster-centric distances splitting the galaxies into centrals and satellites. Their results showed a significantly higher fraction of central quenched galaxies at two virial radii compared to the field centrals, suggesting a non-negligible fraction of quenched, low-mass central galaxies inhabiting the outskirts of massive systems. 

On the other hand, it has been shown that low-mass central galaxies can exhibit strong two-halo conformity, which is the correlation in color or star formation rate (SFR) between central galaxies and their neighbors (e.g., \citealt{Lacerna_2022}, \citealt{Ayromlou_2023}, \citealt{Wang_2023}). \cite{Lacerna_2022} measured the conformity signal using the hydrodynamical simulation IllustrisTNG300 \citep{Weinberger2017,Pillepich_2018} and two catalogs from the semi-analytic model {\sc sag}  applied to the MDPL2 simulation, which differ in the redshift dependence of the SNe feedback efficiency.
They found that low-mass central galaxies in the outskirts of massive systems lead to strong conformity at scales of several megaparsecs at $z=0$. As mentioned above, some environmental effects could be acting beyond the virial radius of galaxy clusters, so we could assume that these effects may contribute to the signal by quenching the galaxies around them. Therefore, a deeper analysis needs to be performed. A recent study performed by \cite{Ayromlou_2023} using the semi-analytical galaxy evolution model, {\sc l-galaxies}, the hydrodynamical simulations IllustrisTNG and EAGLE, and SDSS and DESI observations found that the conformity signal at large scales could be produced by these large-scale environmental effects, such as RPS acting on satellites beyond the halo boundary and on central galaxies in the outskirts of massive halos. They also measured the contribution of former satellite galaxies using the {\sc l-galaxies} semi-analytic model. 
They find that these galaxies contributed between 0\% and 20\% to the conformity signal. On the other hand, using IllustrisTNG300, \cite{Wang_2023} find that former satellites (the term backsplash galaxies is used) can explain the whole signal at $z$ = 0. However, they considered only centrals as neighbor galaxies and  excluded satellites, which modifies the classical definition of the two-halo conformity. 

These studies highlight the importance of the former satellite definition, which can lead to different results. In order to understand the origin and the impact of former satellites on the two-halo conformity signal, we analyzed the galaxy population generated by a hydrodynamical simulation
(IllustrisTNG300) and by a semi-analytic model of galaxy formation {(\sc sag),} to study the evolutionary histories of low-mass central galaxies located in the outskirts of massive clusters or groups.
Then, we analyzed the impact of former satellites on the two-halo conformity at three different epochs, emphasizing the importance of a robust definition of these galaxies to understand their role.
We also report different results when former satellites were accreted by massive and non-massive systems.

The paper is organized as follows. In Sect. 2 we describe the hydrodynamical and semi-analytic data that we used. The samples are described in Sect. 3. Our results on the evolution of low-mass central galaxies are presented in Sect. 4, while the two-halo conformity signal at three epochs is shown in Sect. \ref{sec:conformity}. Our discussions and main conclusions are presented in Sect. \ref{sec:discusion_conclusions}.
%---

%%%%%%%%%%%%%%%%%%%%%%
\section{Data}
%%%%%%%%%%%%%%%%%%%%%%
For the purpose of this work, we  used the suite of hydrodynamical cosmological simulations \textsc{\footnotesize{IllustrisTNG}} and a semi-analytic galaxy catalog \textsc{\footnotesize{MDPL2-SAG}}, built from the galaxy formation model {\sc sag}. The reason for using them is that both datasets revealed a strong two-halo conformity at redshift $z=0$ in low-mass central galaxies near massive systems, making them good models for studying this signal at higher redshifts.
Recent works have also shown that semi-analytic and hydrodynamical models can address different results related to former satellites on conformity (see \citealt{Ayromlou_2023, Wang_2023}). Thus, we expect to contribute to the discussion of previous results with our definition of former satellites in both simulations (see Sect. \ref{section:former_satellites}).

The advantage of using a hydrodynamical model is that its self-consistency allows environmental effects to be applied to all galaxies, whether central or satellite, without the need for prior classification.  
On the other hand, using a semi-analytic model (SAM) of galaxy formation, such as MDPL2-SAG, offers a larger box size, leading to a considerable number of galaxies across all mass ranges, thus improving the statistical analysis. Furthermore, the specific implementation of RPS in {\sc sag}, which affects only satellite galaxies, simplifies the evaluation of how the absence of this process in central galaxies influences the evolution of low-mass galaxies. This characteristic would also enable us to investigate the role of RPS in the strong conformity signal observed in low-mass central galaxies, as reported by \cite{Lacerna_2022}.
Their results also showed similar qualitative trends in the conformity signal for MDPL2-SAG and IllustrisTNG300. 
Therefore, both models provide useful information for understanding the processes these galaxies may have undergone in the past to produce a strong signal at present.

The following subsection provides a detailed description of the datasets employed in this work. Both datasets are summarized in Table \ref{tab:TNG_SAG_table}.\\

%%%%%%
\subsection{IllustrisTNG} 
The IllustrisTNG project (\citealt{Naiman_2018}; \citealt{Nelson_2018}; \citealt{Marinacci_2018}; \citealt{Pillepich_2018}; \citealt{Springel_2018}; \citealt{Nelson_2019}) is a set of cosmological magnetohydrodynamic simulations, TNG50, TNG100, and TNG300, presented in three different volumes (and resolutions), corresponding to 51.7, 110.7, and 302.6 Mpc of box length, respectively. These simulations solve the equations of gravity and magnetohydrodynamics using the moving-mesh code Arepo (\citealt{Springel_2010}). The model includes the contribution of a wide range of physical processes, such as radiative cooling, star formation, supernova and AGN feedback, and chemical enrichment, which drive galaxy formation and evolution. Thus, it is possible to trace the evolution of the gaseous, stellar, and dark matter (DM) components, with results that have been shown to align with observations in the literature
(\citealt{Nelson_2018}; \citealt{Springel_2018}; \citealt{Xu_2019}).

The model assumes a $\Lambda$CDM cosmology with $\Omega_{\Lambda,0} = 0.6911$, $\Omega_{\textrm{m,0}} = 0.3089$, $\Omega_{\textrm{b,0}} = 0.0486$, $\sigma_8 = 0.8159$, $n_{\textrm{s}} = 0.9667$ and $h = 0.6774$ (\citealt{Planck_colaboration_2016}), that follows the evolution of the particles from $z$ = 127 to $z$ = 0. For identifying structures and substructures, the friends-of-friends (FoF) group finder and the \textsc{\footnotesize{SUBFIND}} algorithm (\citealt{Springel_2001}; \citealt{Dolag_2009}) were used in all the snapshots, considering a linking length $b = 0.2$ to identify bound particles. Subhalos with non-zero stellar components are labeled as galaxies. According to our selection criteria for the progenitors, we consider subhalos with stellar masses greater than 50 initial gas cells. In the case of DM particles, 50 particles were used as a threshold. The merger trees were constructed by SubLink (\citealt{Rodriguez_Gomez_2015}) and LHaloTree (\citealt{Springel_2005}): they provide information about the assembly history of galaxies. SubLink and LHaloTree identify all descendant connections belonging to the same FoF group, leading to numerous branches within the galaxy history since each subhalo or galaxy can be linked to one or more progenitors. We have followed galaxies using the main progenitor branch, corresponding to the most massive progenitor identified at each timestep. Access to the data is public and available at \url{www.tng-project.org/}. \\

For the purpose of this work, we %will 
used 
%\textcolor{red}{TNG100} and 
TNG300, corresponding to 
%75 and 
205 $h^{-1}$ Mpc of box length.
%, respectively
The advantage of using 
%TNG100 and 
TNG300 is that 
%they 
it simultaneously follows the evolution of 
%$1820^3$ and 
$2500^3$ %dark matter (
DM particles and gas cells %simultaneously in 
%each 
in the simulation. 
%respectively
This number provides good statistics regarding 
%low-mass galaxies (TNG100) and 
massive structures. \\

\begin{table}
\caption{\label{tab:TNG_SAG_table} Properties of the IllustrisTNG300 simulation and 
\textsc{\footnotesize{MDPL2-SAG}} model.}
\begin{center}
\renewcommand{\arraystretch}{1.6}
\begin{tabular}{l|c|c}
\hline
  & TNG300 & \textsc{\footnotesize{MDPL2-SAG}}\\
 \hline \hline
Volume & (205 $h^{-1}$ Mpc)$^3$ & (1 $h^{-1}$ Gpc)$^{3}$\\
N$_{\rm gas}$ & 2500$^3$ & -\\
N$_{\rm DM}$ & 2500$^3$ & 3840$^{3}$\\
%Mass resolution & - & - & - \\
$m_{\rm baryon}$ & 7.451 $\times$ 10$^{6}$ $h^{-1}$ M$_{\odot}$ & - \\
$m_{\rm DM}$ & 3.996 $\times$ 10$^{7}$ $h^{-1}$ M$_{\odot}$ &  1.5 $\times$ 10$^{9}$ $h^{-1}$ M$_{\odot}$ \\
%Stellar mass cut in log scale [M$_{\odot}$/h] & [9.5, 10] & [9.5, 10]& \\
%Redshift interval & 0 $\leq$ z $\leq$ 2 & 0 $\leq$ z $\leq$ 2 & \\
%N$_{gals}$ & X & 14117 &\\
%N$_{gals}$ Quenched (at z = 0) & & 2255 (16\%)&\\
%N$_{gals}$ SF (at z = 0) & & 11862 (84\%)&\\
%volume evaluating h: 678 Mpc
%mdm evaluating h: 1.017x10^9 Msun
\hline 
\end{tabular}
\end{center}
\end{table}

%%%%%%
\subsection{MDPL2-SAG} 
The \textsc{\footnotesize{MDPL2-SAG}} galaxy catalog was constructed by combining the semi-analytic model of galaxy formation {\sc sag} \citep{Cora_2018} with the cosmological DM \textsc{\footnotesize{MULTIDARK Planck 2}} simulation, \textsc{\footnotesize{MDPL2}} (\citealt{Klypin_2016}; \citealt{Knebe_2018}). The code includes the contribution of several physical processes, such as radiative gas cooling, quiescent star formation, starbursts triggered by mergers and disk instabilities, AGN and SN (Type Ia and II) feedback, and chemical enrichment.
Additionally, the code includes environmental effects like RPS on satellites, with the consequent gradual starvation. The model distinguishes two types of satellites. The first one corresponds to satellite galaxies that maintain their DM halos, whose dynamics allow for tracking galaxy orbits; and the second ones are `orphan galaxies', differentiated from the former because 
the halo finder could not identify their DM halos due to resolution effects. 
We do not consider orphan galaxies in this study.
The DM simulation assumes a $\Lambda$CDM cosmology with $\Omega_\textrm{m} = 0.307$, $\Omega_{\Lambda} = 0.693$, $\Omega_\textrm{B} = 0.048$, $n_{\rm s} = 0.96$ and $H_0 = 100$ $h^{-1}$km s$^{-1}$ Mpc$^{-1}$, where $h = 0.678$ (\citealt{Planck_collaboration_2014}), tracing the evolution of $3840^3$ particles from $z$ = 120 to $z$ = 0 in a box of side length 1 $h^{-1}$ Gpc.

The \textsc{\footnotesize{ROCKSTAR}} halo finder (\citealt{Behroozi_Wechsler_2013}) was used to identify the DM halos and their substructures. This algorithm works with a phase-space based on the standard deviations of the particle distribution in position and velocity space, with a linking length of $b=0.28$, guaranteeing that virial spherical overdensities can be determined for even the most ellipsoidal halos. In this context, any overdensities should comprise at least 20 DM particles. According to the selection criteria, central galaxies are identified as galaxies residing in the center of the potential well of host halos. These main structures can host multiple substructures called subhalos, where satellite galaxies reside. 
For merger trees, \textsc{\scriptsize{CONSISTENTTREES}} (\citealt{Behroozi_2013}) algorithm was used to build up the catalogs.
We follow the evolution of galaxies through their main progenitor branch; that is, the most massive progenitor was identified at each timestep. The \textsc{\footnotesize{MDPL2}} datasets can be found in the \textsc{\footnotesize{CosmoSim}} database\footnote{\url{https://www.cosmosim.org/}}.

\subsection{DM (sub)halo catalogs and fraction of satellites}

Given that the two galaxy catalogs, \textsc{\footnotesize{IllustrisTNG300}} and \textsc{\footnotesize{MDPL2-SAG}}, were constructed using different halo finder algorithms, it is reasonable to expect variations in the fraction of (former) satellite galaxies identified. As we  show later, these discrepancies likely stem from the distinct methodologies employed by \textsc{\footnotesize{SUBFIND}} and \textsc{\footnotesize{ROCKSTAR}}.

\textsc{\footnotesize{SUBFIND}} primarily relies on the particle density field to identify density peaks, assigning the largest one to the main halo. It then analyzes overdensities within the main halo to locate subhalos, excluding unbound particles. This approach focuses solely on spatial distribution.

In contrast, \textsc{\footnotesize{ROCKSTAR}} employs a (sub)structure search that considers the particle positions and velocity fields. This more comprehensive approach can potentially capture a broader range of substructures.

Another factor influencing the results is the linking length ($b$) used in both halo finders. The smaller linking length defined in IllustrisTNG300 might not fully capture the extent of substructures. Conversely, a larger value of the linking length, as defined for \textsc{\footnotesize{MDPL2-SAG}}, could potentially merge distinct but nearby substructures into a single entity.

For both catalogs, we focus on studying how the different properties of these galaxies change over time. We analyze various parameters such as the host halo mass, M$_{\rm 200}$, which represents the mass within a radius R$_{\rm 200}$ where the halo density is 200 times the critical density of the Universe. We also consider the maximum circular velocity (v$_{\rm max}$) representing the highest value of the spherically-averaged rotation curve, whose exact definition changes according to the simulation (see Table \ref{tab:parameters_tng_sag_from_webpages}); the stellar mass (M$_{\rm \star}$) of the subhalo--galaxy, the total gas mass (M$_{\rm gas}$) including both cold and hot components, and the specific star formation rate (sSFR) defined as the ratio of the star formation rate (SFR) 
to the stellar mass (M$_{\rm \star}$). Each parameter description can be found in Table \ref{tab:parameters_tng_sag_from_webpages}.

\begin{table*}
\caption{Parameters taken from the public catalogs from IllustrisTNG and MDPL2 web pages. \label{tab:parameters_tng_sag_from_webpages}}
\begin{center}
\renewcommand{\arraystretch}{1.8}
\renewcommand{\cellalign}{lc}
\begin{tabular}{lclc}
\toprule
Catalog & Parameter & Description & Column name$^*$ \\
\midrule
IllustrisTNG300 & M$_{\rm 200}$ & {\makecell{Total mass of this group enclosed in a sphere \\ whose mean density is 200 times the critical density of \\ the Universe, at the time the halo is considered.}} & Group\_M\_Crit200\\
\addlinespace
 & M$_{\rm \star}$, M$_{\rm gas}$ & {\makecell{Total mass of all member particle/cells\\  which are bound to this Subhalo,\\separated by type. Particle/cells bound to\\subhalos of this Subhalo are NOT accounted for.}} & SubhaloMassType\\
 \addlinespace
% & M$_{gas}$ & {\makecell{Total mass of all member particle/cells\\  which are bound to this Subhalo,\\separated by type. Particle/cells bound to\\subhaloes of this Subhalo are NOT accounted for.}} & SubhaloMassType\\
 \addlinespace
 & SFR & {\makecell{Sum of the individual star formation rates of all\\ gas cells in this subhalo.}} &SubhaloSFR \\ 
 \addlinespace
 & v$_{\rm max}$ & {\makecell{Maximum value of the spherically-averaged rotation \\ curve. All available particle types (e.g., gas, stars, DM,\\ and supermassive black holes) are included\\ in this calculation.}} & SubhaloVmax\\  
\addlinespace
\hline
\addlinespace
MDPL2-SAG & M$_{\rm 200}$ & {\makecell{Mass of the corresponding dark\\  matter halo (M200) in Rockstar.}} & halomass\\
 & M$_{\star}$ & Mass of stars in the spheroid/bulge and the disk. & mstarspheroid + mstardisk\\
 & M$_{\rm coldgas}$ & Mass of gas in the spheroid/bulge and the disk. &  mcoldspheroid + mcolddisk\\ 
 & M$_{\rm hotgas}$ & Hot gas mass. & mhot\\ 
 & SFR & Star formation rate. & sfr\\  
 & v$_{\rm max}$ & {\makecell{Halo circular velocity in physical coordinates,\\ frozen in orphan galaxies.}} & vmax\\   
%MDPL2-SAG & Prototypes (P) & {\makecell{Discriminative\\ Descriptive}} & Nearest \\
\bottomrule
\end{tabular}
\tablefoot{$^*$Based on the column name given in each catalog.}
\end{center}
\end{table*}

%%%%%%%%%%%%%%%%%%%%%%%%%%%%%%
\section{Methodology}
%%%%%%%%%%%%%%%%%%%%%%%%%%%%%%

In this section we describe and define the galaxy selection criteria and the samples used in this study.

%%%%%%%%%%%%%
\subsection{Target and Control samples}
We selected all the low-mass central galaxies in the outskirts of massive systems (M$_{\rm 200} \geq 10^{13}$ $h^{-1}$ M$_{\odot}$) up to 5 $h^{-1}$ Mpc of their centers at $z=0$. 
The range in stellar mass is M$_{\star}$ = [10$^{9.5}$,10$^{10}$] $h^{-1}$ M$_{\odot}$, and it was chosen based on the results obtained by \cite{Lacerna_2022}, where they found a strong conformity signal from these central galaxies at $z = 0$. 
We call these galaxies the target sample.
For TNG300 (MDPL2-SAG), 13,274 (2,495,308) central galaxies at $z = 0$ were identified in the target sample.
Additionally, we chose a control sample with the same stellar mass range as the target sample, except that it was made of central galaxies located farther than 5 $h^{-1}$ Mpc from the centers of galaxy groups and clusters. This selection allowed us to assess whether the proximity to massive systems influences the behavior of low-mass central galaxies. The selection gave us 26,605 (3,814,590) central galaxies in the control sample of TNG300 (MDPL2-SAG). The target and control samples represent the entire sample of central galaxies in that stellar mass range within the catalogs at $z=0$.
For both samples, we follow their evolution until $z = 2$. To identify the TNG300 galaxies in the past, we followed the main progenitor branch for each treeID, which corresponds to the index of the FoF host/parent of each  galaxy.\footnote{The IDs will not be identical in each snapshot.} In other words, we traced their evolution using the SubhaloGrNr parameter. So, if $t_{99}$ corresponds to the snapshot at $z=0$ (snapshot 99 in TNG300), we trace the galaxies backward in time using the previous time steps (e.g., $t_{91}$ for $z=0.1$); this enables us to follow the history of the previous subhalo (i.e., galaxy at snapshot $t_{n-1}$) which shares the same descendant, corresponding to the most massive one derived from LHaloTree and so on.
In the case of MDPL2-SAG, we follow the galaxies in the past using the \texttt{GalaxyStaticID} parameter corresponding to the ID of each galaxy inside the simulation, which will be unique, independent of the simulation snapshot. 
\begin{table}
 \caption{Properties of the target and control samples for IllustrisTNG300.
 \label{tab:target_control_sample_table_TNG300}}
\begin{center}
 %\hspace*{-0.8cm}%
 \renewcommand{\arraystretch}{1.6}
 \begin{tabular}{l|c|c}
 \hline
TNG300 & Target sample & Control sample\\
 \hline \hline
%M$_{\star}$ range in log$_{10}$ scale [M$_{\odot}$/h] & [9.5, 10] & [9.5, 10]\\
 log$_{10}$ M$_{\star}$ [\msun] & [9.5, 10] & [9.5, 10]\\
N$_{\textrm{gals}}$ & 13,274 & 26,605\\
N$_{\textrm{gals}}$ Q (at $z = 0$) & 2,142 (16.1\%) & 2,012 (7.6\%)\\
N$_{\textrm{gals}}$ SF (at $z = 0$) & 11,132 (83.9\%) & 24,593 (92.4\%)\\
 \hline 
 \end{tabular}
\end{center}
\end{table}

\begin{table}
 \caption{Properties of the target and control samples for 
 MDPL2-SAG.}
\label{tab:target_control_sample_SAG}
%\begin{center}
\centering
 %\hspace*{-0.8cm}%
 \renewcommand{\arraystretch}{1.6}
 \begin{tabular}{l|c|c}
 \hline
MDPL2-SAG & Target sample & Control sample\\
 \hline \hline
%Stellar mass range in log$_{10}$ scale [M$_{\odot}$/h] & [9.5, 10] & [9.5, 10]\\
 log$_{10}$  M$_{\star}$ [\msun] & [9.5, 10] & [9.5, 10]\\
N$_{\rm gals}$ & 2,495,308 & 3,814,590\\
N$_{\rm gals}$ Q (at $z = 0$) & 69,718 (2.8\%) & 11,989 (0.3\%)\\
N$_{\rm gals}$ SF (at $z = 0$) & 2,425,590 (97.2\%) & 3,802,601 (99.7\%)\\
 \hline 
 \end{tabular}
%\end{center}
\end{table}

%%%%%%%
\subsection{Separating Q and SF galaxies}

% with Q and SF percentages
\begin{table*}
%\begin{center}
\centering
\caption{Number of primary galaxies at different redshifts. \label{tab:number_primary_galaxies}}
\renewcommand{\arraystretch}{1.6}
\begin{tabular}{l|c|c|c}
\hline
Catalog & $z=0$ & $z=0.3$ & $z=1$\\
 \hline \hline
 IllustrisTNG300 & 39,879 (Q: 10.4\%; SF: 89.6\%) & 41,036 (Q: 2.5\%; SF: 97.5\%) & 37,680 (Q: $<$ 0.1\%; SF: $>$ 99.9\% ) \\ %(Q: 0.05\%, SF: 99.95\% )
 after removing fs & 37,535 (Q: 8.3\%; SF: 91.7\%) & 39,603 (Q: 1.7\%; SF: 98.3\%) & 37,372 (Q: $\ll$ 0.1\%; SF: $>$ 99.9\%) \\ % (Q: 0.003\%, SF: 99.997\%)
 \hline
 MDPL2-SAG & 6,309,898 (Q: 1.3\%; SF: 98.7\%) & 5,309,100 (Q: <1\%; SF: >99\%) & 3,753,659 (Q: <1\%; SF: >99\%)\\
 after removing fs & 6,243,666 (Q: 1.1\%; SF: 98.9\%) & & \\
\hline 
\end{tabular}
\tablefoot{%Number of 
Primary galaxies 
with $M_{\star}$ between $10^{9.5}$ and $10^{10}$ $h^{-1}$ M$_{\odot}$ 
for measuring the two-halo conformity at different $z$. Each panel corresponds to IllustrisTNG300 and MDPL2-SAG galaxy catalogs before and after removing former satellites (fs). The percentages of the primary galaxies in the Q and SF samples are indicated in parentheses.}
%\end{center}
\end{table*}

We separated the samples into quenched (Q) and star-forming (SF) galaxies at $z = 0$ to follow their evolution separately.
To do this, we consider the same sSFR cut used in \cite{Lacerna_2022} to split the samples in both catalogs. In this way, a galaxy is deemed to be quenched if sSFR $\leq$ $10^{-10.5}$ $h$ ${\rm yr}^{-1}$. This cut corresponds to $\sim 10^{-10.7}~ {\rm yr}^{-1}$ with $h = 0.678$ and is chosen because it reproduces well the bimodality of galaxies in the MDPL2-SAG model, as shown in previous studies such as \cite{Brown_2017} and \cite{Cora_2018}. 
We apply the same sSFR cut in both catalogs for a direct comparison, avoiding biased results that could arise from using different cuts. An explicit comparison of the sSFR distribution as a function of the stellar mass in TNG300 and MDPL2-SAG was shown in \citet[][see their Fig. 1]{Lacerna_2022}. 
We checked that using a different cut in sSFR of $10^{-10.8}$ $h$ ${\rm yr}^{-1}$, which corresponds to $\sim 10^{−11}~ {\rm yr}^{-1}$  with $h = 0.678$, 
the overall evolution of Q or SF low-mass central galaxies does not change.

From the target sample in TNG300, we find 2,142 ($\sim$ 16\%) Q galaxies and 11,132 ($\sim$ 84\%) SF galaxies at $z = 0$, while for the control sample, we find 2,012 ($\sim$ 8\%) Q and 24,593 ($\sim$ 92\%) SF galaxies. Table \ref{tab:target_control_sample_table_TNG300} gives all the information about both samples in IllustrisTNG300. 
We find a fraction significantly greater of Q galaxies for the target sample than in the control sample. For SF galaxies, in both cases, the percentages are greater than 80\%, even for the target sample, suggesting that most of these galaxies are still forming new stars in environments close to massive systems.
The median stellar mass of Q galaxies is log$_{10}$(M$_{\rm \star}$/\msun) = 9.74 and 9.77 for target and control samples, respectively, whose 16th and 84th percentiles range between log$_{10}$(M$_{\rm \star}$/\msun) = 9.58 and 9.91 for the target sample, and between 9.59 and 9.94 for the control galaxies. Therefore, the target and control samples of Q central galaxies are well-mass-matched. The same occurs with the SF galaxies despite the number of control galaxies being larger than the target sample. Their median stellar mass is log$_{10}$(M$_{\rm \star}$/\msun) = 9.73 in both samples.

From the target sample in MDPL2-SAG, we find 69,718 ($\sim$ 3\%) Q galaxies and 2,425,590 ($\sim$ 97\%) SF galaxies, while for the control sample, we find 11,989 ($\ll$ 1\%) Q and 3,802,601 ($>$ 99 \%) SF galaxies. The differences in number are given by the box size of the semi-analytic model, which is significantly greater than TNG300 and, therefore, gives us better statistics in the mass range of this study. It is important to note that for the semi-analytic model, the percentages for Q galaxies are considerably lower compared to the fractions found in IllustrisTNG. 
Nonetheless, we find for the semi-analytic model a behavior similar to that in TNG300, where the number of SF low-mass central galaxies is greater than the number of Q galaxies, which in the case of MDPL2-SAG increases to 97\% for the target sample. 
Table \ref{tab:target_control_sample_SAG} summarizes the properties of the target and control samples for MDPL2-SAG.
The median stellar mass of Q galaxies is log$_{10}$(M$_{\rm \star}$/\msun) = 9.72 and 9.76 for target and control samples, respectively, whose 16th and 84th percentiles range between log$_{10}$(M$_{\rm \star}$/\msun) = 9.56 and 9.91 for the target sample, and between 9.57 and 9.96 for the control galaxies. Therefore, both samples of Q central galaxies are well mass-matched despite the number of target objects being larger than the number of control galaxies. In the case of SF galaxies, although the number of control galaxies is larger than the number of target objects, the samples are well-mass-matched. Their median stellar mass is log$_{10}$(M$_{\rm \star}$/\msun) = 9.72 in both samples, whose 16th and 84th percentiles range between log$_{10}$(M$_{\rm \star}$/\msun) = 9.57 and 9.90 for the target sample and between 9.56 and 9.90 for the control galaxies.

The numbers in Table \ref{tab:target_control_sample_SAG} support the hypotheses that there are more SF galaxies in the low-mass regime than Q galaxies.
The Q fraction of galaxies increases when low-mass central galaxies are close to galaxy groups and clusters, but the fraction of SF galaxies still dominates.

%%%%%%%%%
\subsection{Former satellites} \label{section:former_satellites}
Since there is a chance that the main progenitor of a central galaxy at present might have crossed the virial radius of another system in the past, which may produce contamination in our samples of central galaxies, we have identified former satellites (fs) in each catalog. Given that TNG300 and MDPL2-SAG are different built galaxy catalogs,
we have applied different criteria to select the former satellites (possible polluting galaxies). 
These definitions were applied to enhance the comprehension of the role of these galaxies in the signal of the two-halo conformity, mostly because previous work showed controversial results in this regard (see Sect. \ref{sec:intro}). 

For TNG300, we consider the information provided by the subhalo finder for the main progenitors of target and control galaxies and add a distance criterion. 
We select the satellite candidates identified as ``satellite'' by the subhalo finder in one or more snapshots at $z > 0$ and measure the distance between the satellite candidate and their respective central galaxy according to the subhalo finder at that time. If the distance satisfies $d \leq$ 2R$_{200}$, the progenitor galaxy is considered a satellite in that snapshot, that is, a former satellite. Otherwise, the progenitor is regarded as a central galaxy.

For MDPL2-SAG, we also considered the information provided by the subhalo finder, and we paid attention to the existence of one more member within the respective ({\sc rockstar}) halo in that snapshot. We realized that there were cases where the main progenitor of a central galaxy at present was flagged as a satellite in some snapshot(s) in the past by the halo finder, but there were no other galaxies in the same host halo (in that snapshot). Therefore, if the progenitor galaxy is a satellite according to the subhalo finder but the only one galaxy in the host halo, then the galaxy is still considered as central by us. Otherwise, the galaxy will be classified as a former satellite if other galaxies (central, satellites within a DM substructure, or orphans) are in the same halo. 
For the two-halo conformity at higher redshifts (Sect. \ref{subsec:conformity_high_z}), we have applied the same criteria (per model) to select former satellites, that is, selecting candidates that satisfy all the conditions mentioned above for the progenitors of primary galaxies at $z = 0.3$ and $z = 1$.

%%%%%%%%%
\subsection{Primary and secondary samples for the analysis of conformity}
\label{sebsec_primary_secondary}

Because we are also interested in studying the conformity signal (Sect. \ref{sec:conformity}), we have defined two other samples: primary and secondary samples. The idea is to analyze the correlation in the sSFR of low-mass central galaxies (primary galaxies) with their neighbors (secondary galaxies). The primary sample at $z=0$ corresponds to both target and control galaxies, and the secondary sample is all the neighboring galaxies (either central or satellite) that inhabit within a radius of 10 $h^{-1}$ Mpc from the primary galaxies. Some authors argue that the inclusion of satellites in the secondary sample for the measurement of conformity can produce some contamination in the results, mainly because some surveys cannot distinguish properly between a central or satellite (\citealt{Kauffmann_2013, Sin_2017, Tinker_2017, Lacerna_2018}). However, satellite neighbors also play an important role in the signal measured. Because we are using simulated data, in this study, we consider both central and satellite as secondary galaxies with stellar masses greater than $10^{9}$ $h^{-1}$ M$_{\odot}$. This lower limit in the stellar mass has been used to avoid resolution effects in previous works (e.g., IllustrisTNG300: \citealt{Pillepich_2018a}, \citealt{Montero-Dorta_2020}, \citealt{Lacerna_2022}; MDPL2-SAG: \citealt{Haggar_2020}, \citealt{Hough_2023}). We applied this criterion in both simulations when measuring the two-halo conformity in Sect. \ref{sec:conformity}.

We measured the mean quenched fraction ($f_{\rm Q}$) of neighboring (secondary) galaxies as a function of the distance from the primary samples (Q and SF primaries) to 10 $h^{-1}$ Mpc. 
The quenched fraction of neighboring galaxies is measured at different redshifts. We perform the analysis up to $z = 1$ because of the low number of Q central galaxies in the stellar mass range of $10^{9.5}$ to $10^{10}$ $h^{-1}$ M$_{\odot}$ at higher redshifts.
Table \ref{tab:number_primary_galaxies} summarizes the properties of all the primary samples at three different redshifts ($z$ = 0, 0.3, and 1) for both IllustrisTNG300 and MDPL2-SAG galaxy catalogs.

%%%%%%%%%%%%%%%%%%%

%%%%%%%%%%%%%%%%%%%%%%%%%%%%%%%%%%%%%%%%%%%%%%%%
\section{The evolution of low-mass central galaxies}

\begin{figure*}
  \centerline{\includegraphics[scale=0.195]{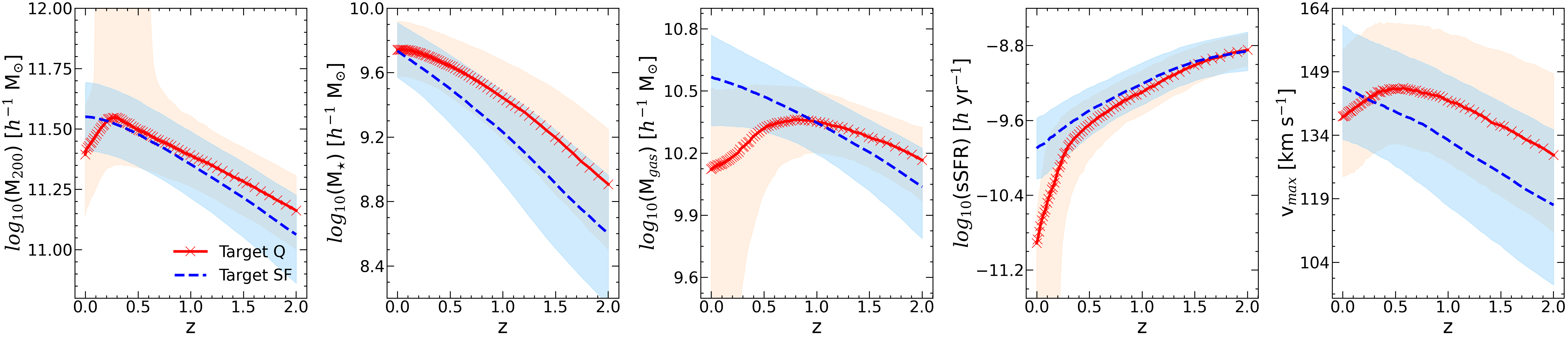}}
  \centerline{\includegraphics[scale=0.195]{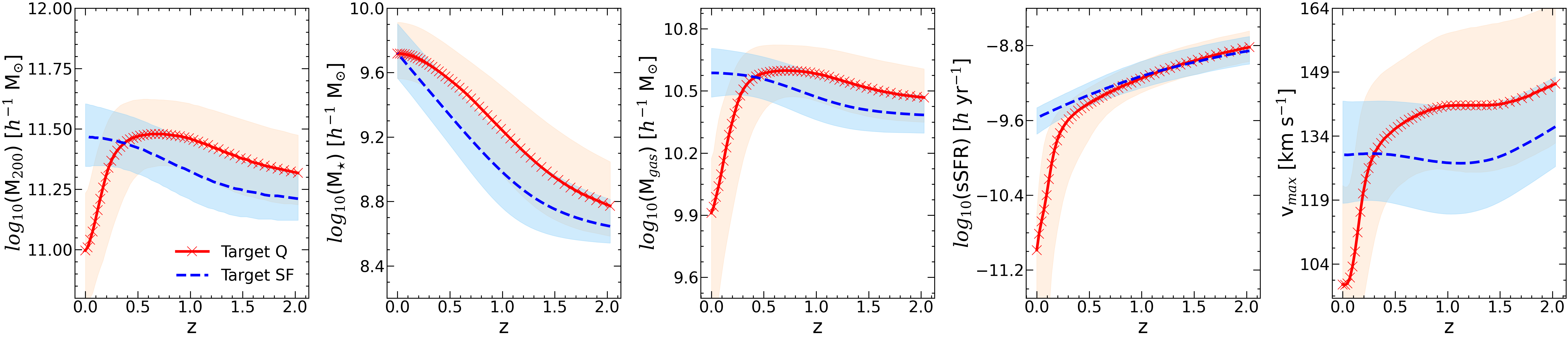}}
\caption{Median trends of the target sample from TNG300 (top) and  MDPL2-SAG (bottom). From left to right: M$_{\rm 200}$($z$), M$_{\rm \star}$($z$), M$_{\rm gas}$($z$), sSFR($z$), and v$_{\rm max}$($z$). The red and blue lines represent the Q and SF samples, respectively. The shaded regions indicate the 16th–84th percentile ranges. \label{fig:all_params_target_TNG_and_SAG}}
\end{figure*}

Our interest lies in studying the large-scale environmental effects on low-mass central galaxies located in the outskirts of groups and clusters. These effects may lead to their cessation of SF activity. Additionally, we aim to explore their potential contribution to the conformity signal at large scales. To achieve this, we present the evolutionary histories of these galaxies from the TNG300 and MDPL2-SAG datasets. Figure \ref{fig:all_params_target_TNG_and_SAG} shows the evolution of the median trends of the host halo mass, M$_{\rm 200}(z)$, the stellar mass, M$_{\rm \star}(z)$, the total gas content, M$_{\rm gas}(z)$, the specific star formation rate, sSFR$(z)$, and the maximum circular velocity, v$_{\rm max}(z)$ from left to right for TNG300 (upper panel) and MDPL2-SAG (lower panel) catalogs.
Each panel shows the evolution of Q (red) and SF (blue) target galaxies. The shaded contours correspond to the 16th and 84th percentiles for each sample. In addition, Figs. \ref{fig:target_control_q_sf_TNG} and \ref{fig:target_control_q_sf_SAG} show the same as in Fig. \ref{fig:all_params_target_TNG_and_SAG} but separate the Q (upper) and SF (lower) galaxies to compare the histories with the control sample, which is shown in gray. The following section will describe the analysis performed for each parameter of interest.

%%%
\subsection{Analysis of physical parameters}
\label{subsec_evolution}

\subsubsection{Host halo mass growth}

% TNG300 
\begin{figure*}
\centering
\subfigure{\includegraphics[scale=0.195]{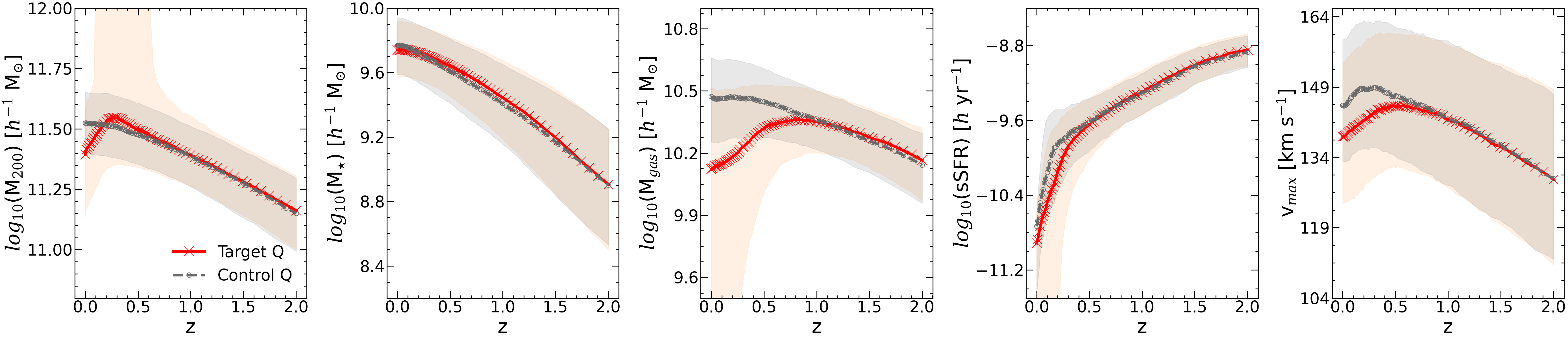}} %\par\smallskip
\\[-1.7ex]
\subfigure{\includegraphics[scale=0.195]{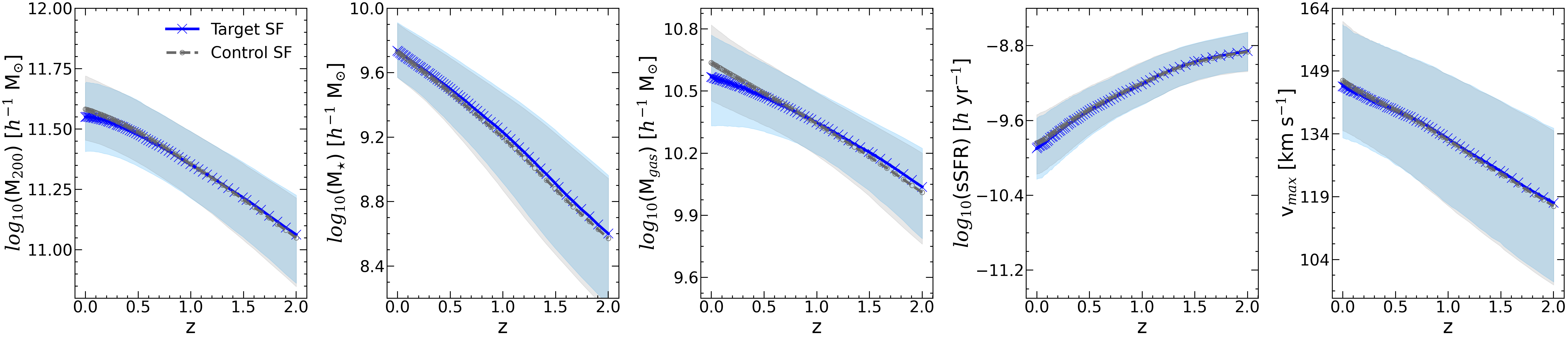}}
\caption{Same parameters as Fig. \ref{fig:all_params_target_TNG_and_SAG}. The upper panels show the median trends for Q galaxies (target sample in red, control sample in gray), and the lower panels show the SF galaxies for the target (blue) and control (gray) samples in TNG300.
\label{fig:target_control_q_sf_TNG}}
\end{figure*}

% SAG
\begin{figure*}
\centering
\subfigure{\includegraphics[scale=0.195]{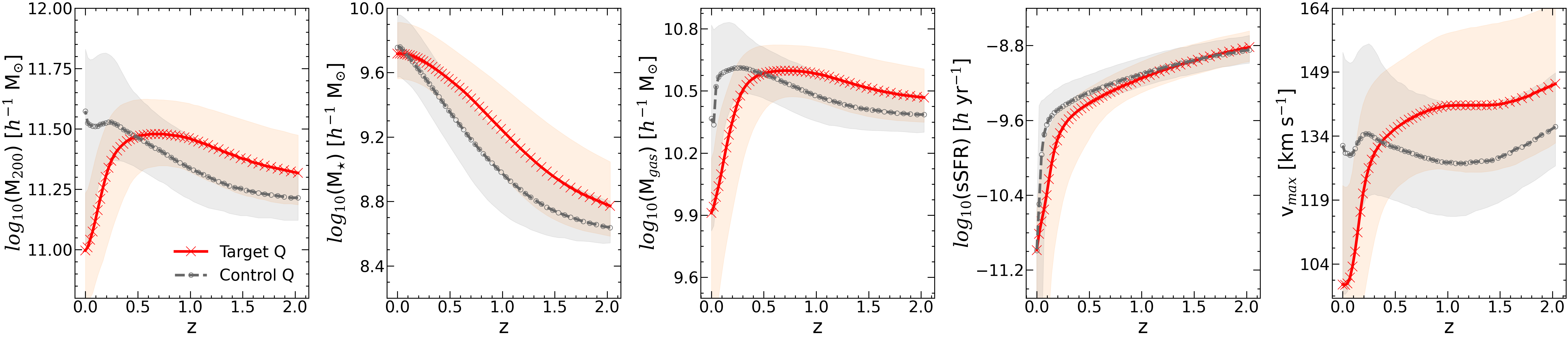}} %\par\smallskip
\\[-1.7ex]
\subfigure{\includegraphics[scale=0.195]{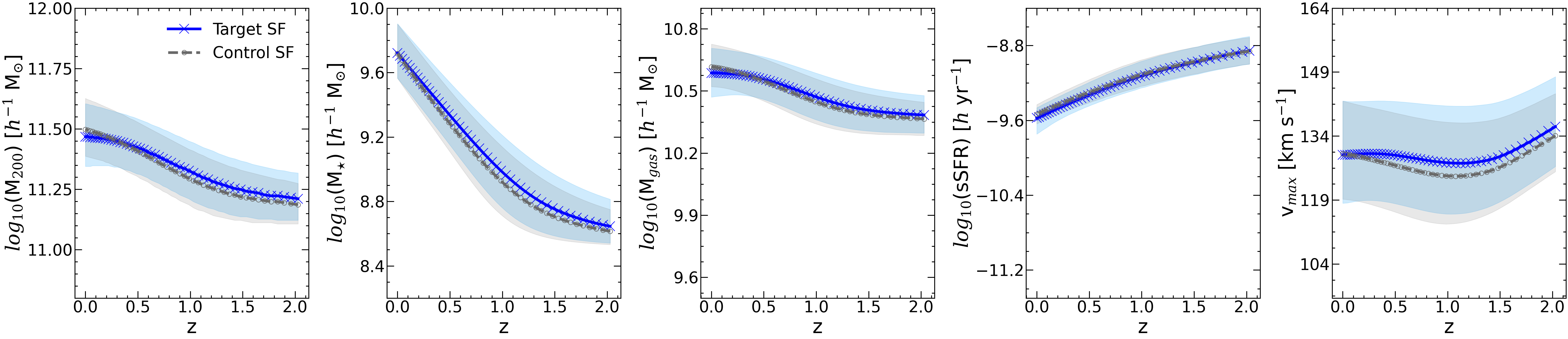}}
\caption{Same parameters as Fig. \ref{fig:all_params_target_TNG_and_SAG}. The upper panels show the median trends for Q galaxies (target sample in red, control sample in gray), and the lower panels show the SF galaxies for the target (blue) and control (gray) samples in MDPL2-SAG.\label{fig:target_control_q_sf_SAG}}
\end{figure*}

One of our hypotheses regarding the star formation quenching of these low-mass central galaxies near massive halos involves the influence on their growth. To verify the existence of this effect, we examine the growth of the host halo mass of target and control galaxies. The left panel of Fig. \ref{fig:all_params_target_TNG_and_SAG} shows the evolution of the median M$_{\rm 200}(z)$ for Q (red line) and SF (blue line) target galaxies. In addition, we have added the v$_{\rm max}$ parameter (halo circular velocity, right panel) as a dynamic mass proxy. For TNG300, SF galaxies were hosted by less massive halos than Q galaxies, at least down to $z \sim 0.3$, where the lines converge mostly to the same median values. At this redshift, there is a considerable fraction of Q galaxies that inhabited more massive halos (see the shaded region), with masses above 10$^{13}$ $h^{-1}$ M$_{\odot}$ (its 84th percentile is about 10$^{14}$ $h^{-1}$ M$_{\odot}$).
This trend was not observed for SF galaxies. At lower redshifts, there is a steep drop in halo growth for the Q target galaxies, reaching a median value of $\log_{10}$(M$_{\rm 200}$/\msun) = 11.394 at $z=0$. 
To see if this behavior is exclusive to Q galaxies near massive halos, we compare it with the M$_{\rm 200}(z)$ of Q control galaxies.
We see in Fig. \ref{fig:target_control_q_sf_TNG} that the host halo mass growth for the control sample is quite different than the Q targets at lower redshifts, reaching a higher median value of $\log_{10}$(M$_{200}$/\msun) = 11.524 at $z=0$. The growth of the halo mass in SF galaxies evolves monotonically in both target and control samples and resembles the trend observed in Q control galaxies. The median value for SF galaxies in the target sample reaches $\log_{10}$(M$_{\rm 200}$/\msun) = 11.549 at $z=0$, being slightly lower than SF control galaxies, which reach a median value of $\log_{10}$(M$_{200}$/\msun) = 11.582.

On the other hand, for MDPL2-SAG we find that SF galaxies evolve inhabiting less massive halos than their Q counterpart, at least until $z$ $\sim$ 0.3 (same behavior found in TNG300). At lower redshifts, the median trend for Q target galaxies drops dramatically, increasing the differences between the host halos from Q and SF target galaxies at $z = 0$, with median values of $\log_{10}$(M$_{\rm 200}$/\msun) = 10.997 and 11.468, respectively. Figure \ref{fig:target_control_q_sf_SAG} shows that the rapid drop in the host halo mass occurs only for target galaxies (i.e., those inhabiting the outskirts of galaxy groups and clusters). 
The median trend for control galaxies in MDPL2-SAG  is qualitatively similar to those found for control galaxies in TNG300. For Q control galaxies, the median value at $z=0$ is $\log_{10}$(M$_{\rm 200}$/\msun) = 11.574.
In the case of SF galaxies, small differences are found at $z > 0.5$, where target galaxies are hosted in slightly more massive DM halos than SF control objects. And when reaching $z=0$ the trend is reversed, with a median value of $\log_{10}$(M$_{200}$/\msun) = 11.498 for control galaxies.

The analysis shows that there is a drop in the host halo mass at $z \lesssim 0.3$ for Q galaxies that currently reside near massive halos in both TNG300 and MDPL2-SAG. This behavior is not observed in Q central galaxies of the same stellar mass located farther from massive groups and clusters. It is tempting to speculate that the halo mass drop is behind the quenching of target galaxies that, in turn, produce the two-halo conformity.
An interesting difference found between both models is that, for MDPL2-SAG, we have not found target galaxies that would have been hosted by massive halos with masses M$_{\rm 200}$ $\geq$ 10$^{13}$ $h^{-1}$ M$_{\odot}$ in the past. We study in more detail the role of former satellite galaxies in the evolution of the Q target galaxies below in Sect. \ref{subsec:removing_fs}.

In the case of v$_{\rm max}$, the Q target galaxies in TNG300 show a peak around $z \sim 0.5$ followed by a drop (similar as found for M$_{\rm 200}$). This peak does not occur for MDPL2-SAG. Instead, v$_{\rm max}$ slightly decreases until $z \sim 0.5$, followed by a steep drop in their median values. The discrepancies at $z > 0.5$ may be attributed to how the v$_{\rm max}$ is calculated in each catalog. In the case of IllustrisTNG300, the v$_{\rm max}$ is computed at the subhalo scale using dark matter and baryonic particles, whereas, for MDPL2-SAG, the v$_{\rm max}$ parameter is measured using only dark matter particles enclosed within R$_{200}$. 
Despite these differences, this parameter can allow us to gain insight into the dynamic state of the host (sub)halos in each catalog. Both models, in turn, show a drop in v$_{\rm max}$ at $z \lesssim 0.5$ in Q target galaxies, which is consistent with their drop in halo mass at lower redshifts and quenched control galaxies with larger v$_{\rm max}$ values than Q targets at present (see upper-right panels in Figs. \ref{fig:target_control_q_sf_TNG} and \ref{fig:target_control_q_sf_SAG}). 
A difference with respect to halo mass is that quenched control galaxies also exhibit a drop in v$_{\rm max}$. Still, it occurs at lower redshift and with a smaller amplitude than that for Q target galaxies.
In the case of SF target galaxies, each model shows evolutionary histories comparable to their control samples, showing no significant differences concerning the location of these galaxies with respect to groups or clusters. 
\subsubsection{Evolution of the gas and stellar mass content}
The second and third panels of Fig. \ref{fig:all_params_target_TNG_and_SAG} show the evolution of the stellar mass and total gas content, respectively. 

The median trends of target galaxies in TNG300 and MDPL2-SAG are quite similar for stellar mass. Q galaxies exhibit a higher stellar mass over the $z$ range than SF galaxies. If we compare their median trends with the control samples (see Figs. \ref{fig:target_control_q_sf_TNG} and \ref{fig:target_control_q_sf_SAG}), the evolutionary paths are almost indistinguishable for TNG300.
In the case of MDPL2-SAG, some distinction can be seen during the stellar mass evolution, being Q target galaxies slightly richer in stellar mass than Q control galaxies (see upper panel of Fig. \ref{fig:target_control_q_sf_SAG}). The median trends for SF target and control (see lower panel of Fig. \ref{fig:target_control_q_sf_SAG}) remain pretty similar as in TNG300.

For the evolution of the gas content, the Q target galaxies in TNG300 reach a maximum around $z \sim$ 1 and then decrease their amount until $z = 0$. The 16th percentile and median comparison with the control sample (Fig. \ref{fig:target_control_q_sf_TNG}) show that the gas content loss becomes more important for those galaxies in the outskirts of massive groups and clusters, increasing this gap at $z = 0$ whose median values for the target galaxies are log$_{10}$(M$_{\rm gas}$/\msun) = 10.122 and 10.568 for Q and SF galaxies. In the case of control galaxies, the median values reach log$_{10}$(M$_{\rm gas}$/\msun) = 10.473 and 10.636 for Q and SF galaxies, respectively. The different behavior in  M$_{\rm gas}$($z$) for Q galaxies in the target and control samples at $z < 1$ resembles the drop in M$_{\rm 200}$ for the target galaxies at $z \lesssim 0.3$, which suggests the quenching mechanism of target galaxies is different than for the control sample in TNG300.

The scenario in MDPL2-SAG, on the other hand, shows that Q target galaxies experienced more gas loss than their SF counterpart, which drives the quenching of these galaxies at $z = 0$. The drop in gas mass is even more dramatic compared to TNG300, a similar behavior as found for the host halo mass in Q galaxies. This decrease in gas mass behaves similarly to the findings by \cite{Salerno_2022}, where galaxies in infall and filamentary regions exhibit a drop in the mass of hot gas from $z \sim 0.5$ to the present. The median values in the gas mass at $z=0$ for the target galaxies are log$_{10}$(M$_{\rm gas}$/\msun) = 9.911 and 10.587 for Q and SF galaxies. If we compare the median trends with the control sample, no major differences are found for SF galaxies. Instead, Q control galaxies exhibit a slight drop in the last snapshots, which was not observed in TNG300. The median values for control galaxies reach log$_{10}$(M$_{\rm gas}$/\msun) = 10.367 and 10.616 for Q and SF galaxies.

As a summary of the gas mass, both TNG300 and MDPL2-SAG galaxy catalogs qualitatively follow the same trend in the evolution of galaxies inhabiting near massive groups and clusters. For the control galaxies, small differences arise mainly in the gas mass for the Q galaxies at $z \lesssim 0.1$ because of a small decrease in MDPL2-SAG. The latter might be connected to the slight drop in v$_{\rm max}$ and M$_{\rm 200}$ for these galaxies.

\subsubsection{sSFR history}
Since low-mass central galaxies in the vicinity of massive systems appear to be more quenched than those galaxies inhabiting regions farther than 5 $h^{-1}$ Mpc at $z = 0$, we follow their sSFR history to estimate if these differences were milder in the past.
The fourth panel (from left to right) of Fig. \ref{fig:all_params_target_TNG_and_SAG} shows the median sSFR history of both Q and SF galaxies. Both synthetic catalogs show a decreasing evolutionary trend until $z = 0$. At higher $z$, Q and SF samples have very similar median values of sSFR.
The evolutionary paths diverge at $z \lesssim$ 1.5 in TNG300 (upper panel) and $z \lesssim$ 1 in SAG (lower panel), where the Q galaxies exhibit a steeper drop than the SF counterpart. We can notice from both models that this drop in sSFR of Q galaxies is independent of the distance of these galaxies to a group or cluster, which can be seen clearly in Figs. \ref{fig:target_control_q_sf_TNG} and \ref{fig:target_control_q_sf_SAG}, where Q target galaxies are compared with Q control galaxies.

Interestingly, the slope of Q galaxies in both simulations becomes steeper at $z \sim$ 0.3, dropping to sSFR $\sim$ $10^{-11}$ $h$ yr$^{-1}$ at $z = 0$. This behavior might be a consequence of the decrease in the host halo mass growth, v$_{\rm max}$, and gas mass previously found for these galaxies.

Given that the target and control samples have stellar masses between log$_{10}$(M$_{\star}$/\msun) = [9.5, 10] at $z=0$, we are aware that there might be resolution problems at higher redshifts. However, we have shown that our important results occur at $z < 1$ where typically the progenitor stellar masses are $>10^9$ $h^{-1}$ M$_{\odot}$, so that resolution effects should not affect the analysis of the evolution of these galaxies.

The results presented in this section for both IllustrisTNG300 and MDPL2-SAG catalogs showed comparable histories for SF galaxies, with no particular distinction between target and control galaxies. However, for Q galaxies, their evolution shows significant drops in their halo mass, gas mass, v$_{\rm max}$, and sSFR, which are pronounced for galaxies close to massive systems.

For the Q target galaxies in IllustrisTNG300, the drop in the gas mass occurs at $z \lesssim 0.5$, suggesting that galaxies could have suffered  RPS or starvation. This scenario is supported by the large number of galaxies that started to inhabit more massive halos at this epoch (see 84th percentile for M$_{200}$ in Fig.~\ref{fig:target_control_q_sf_TNG}, for instance), affecting their sSFR later. The gas mass loss causes a subsequent drop in their v$_{\rm max}$, given how it is calculated (including the baryonic component). A final drop appears in M$_{200}$ around $z \lesssim 0.3$, probably due to the slow action of tidal stripping (TS) exerted by the larger halos, gradually disrupting the host DM halo of target galaxies. The decrease in M$_{200}$ is not observed for the control galaxies, which could be related to the fact that TS did not happen because these galaxies were typically located far from massive structures.
The progenitors of central galaxies that were part of larger halos as satellites may pollute the trends in the evolution of target galaxies. The analysis after removing those ``polluting'' former satellite galaxies is shown below in Sect.~\ref{subsec:removing_fs}.

For MDPL2-SAG, Q target galaxies exhibited a simultaneous drop in M$_{200}$ and v$_{\rm max}$. As control galaxies do not show such a drop, the observed decrease in the target galaxies is likely due to their proximity to massive systems. This, in turn, points to the presence of strong gravitational interactions within the model for galaxies inhabiting the outskirts of massive systems. Interestingly, the progenitors of these galaxies were never part of a massive system, that is, they were not exposed to the effects of RPS. 
It is possible that the decrease in M$_{200}$ and v$_{\rm max}$ may be related to the high-density environment around massive structures, such as filaments. These strong interactions also seem to affect their gas mass, as we observe a drop in M$_{\rm gas}$ almost simultaneously with the drop in halo mass. The outcome is a drop in their sSFR about 370 Myr later.

%%%%%%%%%%%%
\subsection{Removing former satellites} 
\label{subsec:removing_fs}

%TNG
\begin{figure}
  \centerline{\includegraphics[width=8.8cm, height=4cm]
  {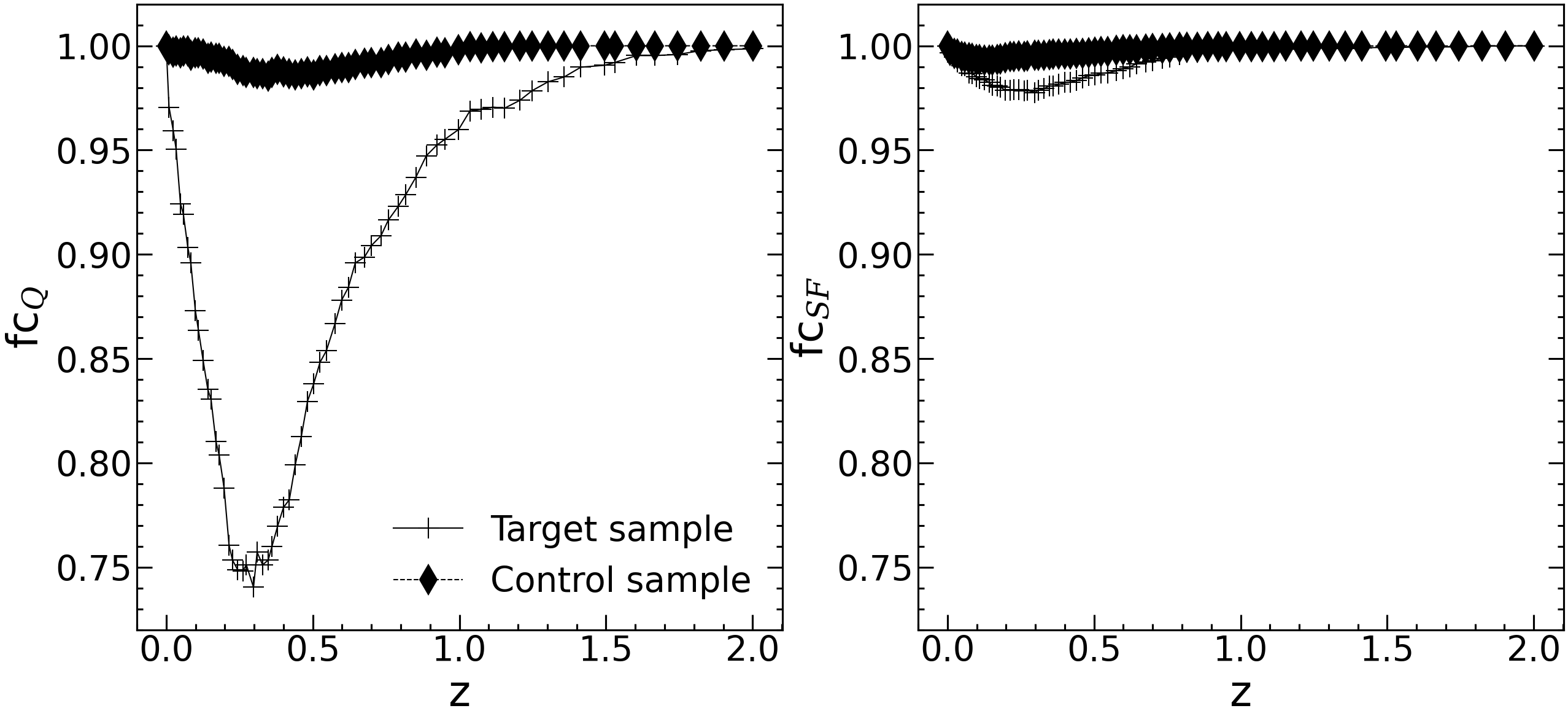}}
  \caption{Evolution in the fraction of central galaxies for Q (left) and
SF (right) galaxies of the target and control samples in TNG300 (crosses and diamonds, respectively). The fraction is equal to unity at $z=0$ by definition.\label{fig:fraction_former_satellites_TNG}}
\end{figure}

%SAG
\begin{figure}
  \centerline{\includegraphics[width=8.8cm, height=4cm]{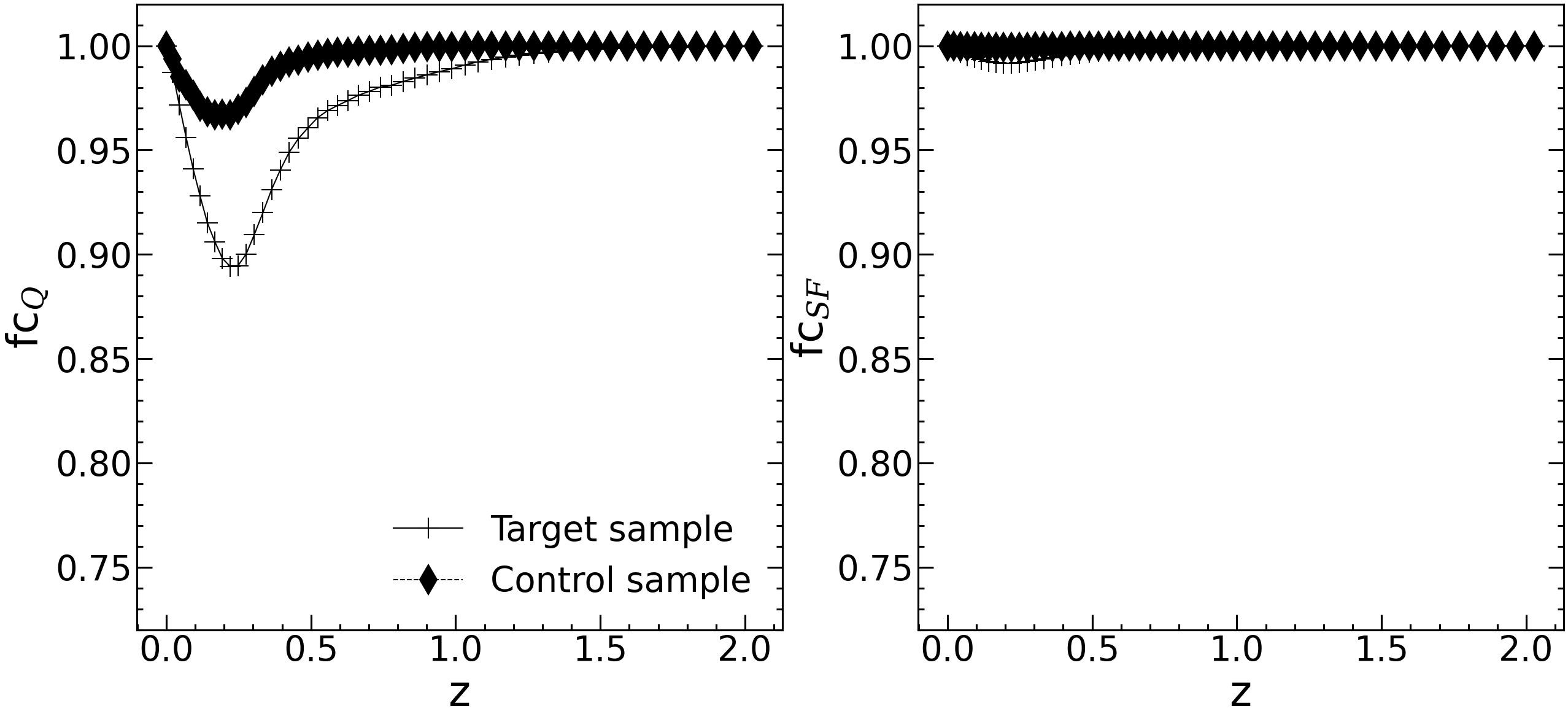}}
  \caption{Same as Fig. \ref{fig:fraction_former_satellites_TNG}, but for the target and control samples in MDPL2-SAG.\label{fig:fraction_former_satellites_SAG}}
\end{figure}

As previous studies have shown (e.g., \citealt{Ayromlou_2023} and \citealt{Wang_2023}), it has been demonstrated that former satellites contribute to the two-halo conformity signal, but with controversial results about their impact. The latter highlights the importance of suitable sampling of former satellites (we detail the definitions used in this study in Sect. \ref{section:former_satellites}). 
The results in the previous section on the halo mass growth showed that massive DM halos hosted a significant fraction of Q target galaxies at $z \sim 0.3$ in TNG300.
This phenomenon indicates that a fraction of low-mass central galaxies may have previously been satellites of more massive systems, influencing their properties. Therefore, it is plausible that environmental effects acted on those galaxies during that time.
In order to assess this effect, we first analyzed how the fraction of low-mass central galaxies (f$_c$) changed in the past for both TNG300 and MDPL2-SAG catalogs. By definition, f$_c$ = 1 at $z=0$; if their progenitors were always central galaxies, then f$_c$ = 1 at all redshifts. Figures \ref{fig:fraction_former_satellites_TNG} and \ref{fig:fraction_former_satellites_SAG} show how the fraction of central galaxies for the target (crosses) and control (diamonds) samples varied at different epochs during their evolution in each catalog. 
As we can see in the figures, a significant fraction of Q target galaxies (left panels) were satellites at some time in the past, reaching a maximum at $z \sim 0.3$ in both models, with about 25\% of the targets for the hydrodynamical model and 11\% for the semi-analytic model at that time. 
The identification of former satellites is done in halos of any mass, but these results elucidate the observed trend in TNG300 in which massive DM halos hosted a significant fraction of Q target galaxies. 
In the case of Q control galaxies, the percentages reach around 2\% and 4\% at that epoch for TNG300 and MDPL2-SAG, respectively, indicating it is more likely that galaxies have been part of other hosts in the past if they are inhabiting regions close to massive groups or clusters. For SF galaxies, the percentages of central galaxies that were satellites in the past are quite low compared to the Q sample, which decreases even more if the galaxy belongs to the control sample.

%TNG
\begin{figure*}
\centering{\includegraphics[width=150mm]{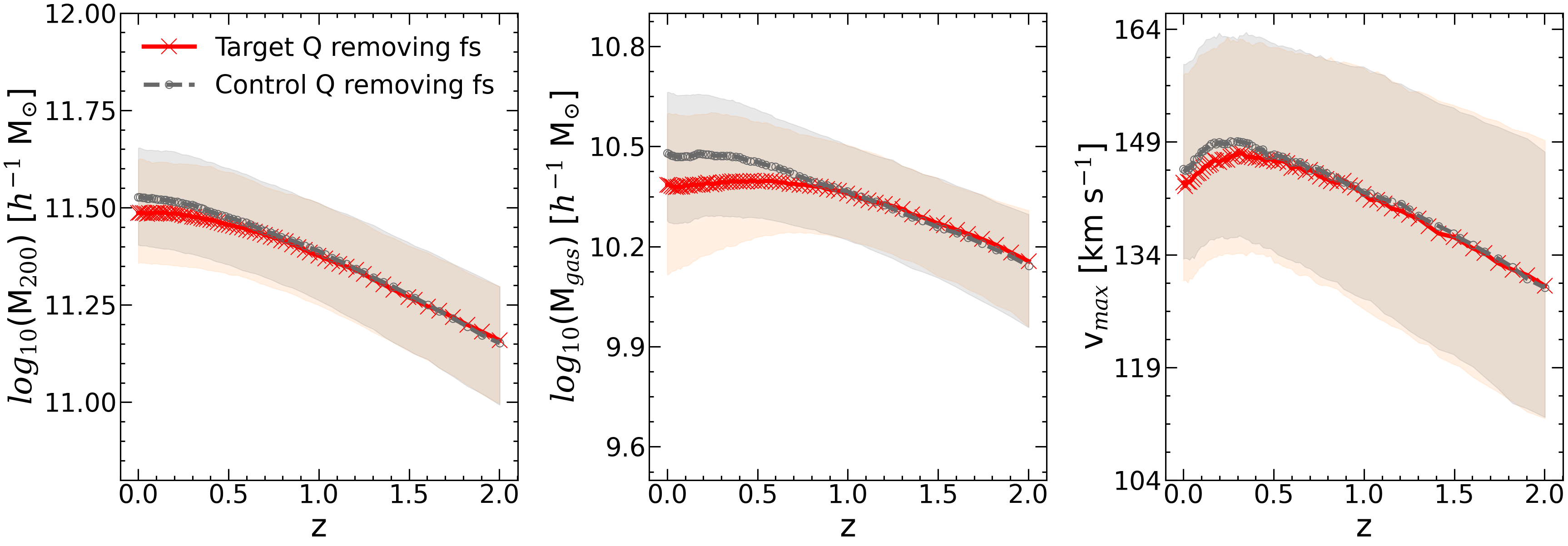}} %\par\smallskip
%\subfigure{\includegraphics[width=180mm]{results/target_control_SF_removing_backsplash.png}}
\caption{Evolution of the host halo mass (left), gas mass (middle), and $v_{max}$ (right) for the target and control Q galaxies, after removing former satellites in TNG300.
\label{fig:target_control_q_removing_fs_TNG}}
\end{figure*}

The cumulative percentages of former satellites at $z=0$ found for TNG300 are 45\% (Q) and 7\% (SF) for target galaxies and 4\% (Q) and 2\% (SF) for the control sample. Instead, for MDPL2-SAG, 17\% (Q) and 2\% (SF) were satellites for the target sample, and 6\% (Q) and less than 1\% (SF) for the control galaxies. All this information is summarized in Table \ref{tab:number_target_control_former_satellites_TNG_SAG}. 

The differences in the fraction of former satellites between the models are related to the numerical resolution of each simulation. Since MDPL2-SAG has a lower resolution than TNG300, it would be expected to have a lower fraction of former satellites. To confirm this, we performed a test by calculating the fraction of former satellites using other simulations from the TNG suite with different resolutions. We used the TNG300-2 with a lower DM mass resolution of $m_{\rm {DM}} =$ $3.2 \times 10^8$
$h^{-1}$ M$_{\odot}$. We found 31.4\% of former satellites in the target Q sample, lower than the 45\% reported for TNG300-1. We then repeated the analysis, this time using a simulation with better resolution, TNG100-1 with $m_{\rm {DM}} = 5.1 \times 10^{6}$ $h^{-1}$ M$_{\odot}$, and we found that 73\% of the target Q sample were identified as former satellites. \\

\begin{table}
\caption{Number of former satellites, $N_{\rm fs}$, in each sample.\label{tab:number_target_control_former_satellites_TNG_SAG}}
%\begin{center}
\centering
\renewcommand{\arraystretch}{1.6}
\begin{tabular}{|l c c|c c|}
\hline
 & TNG300 & & SAG & \\
\hline \hline
 & $N_{\rm fs}$ & \% &  $N_{\rm fs}$ & \% \\
 \hline \hline
Target Q & 967 & 45.1\% &  11,899 & 17.1\%\\
Target SF & 766 & 6.9\% & 46,663 & 1.9 \%\\
\hline
Control Q & 72 & 3.6\% & 678 & 5.7\%\\
Control SF & 539 & 2.2\% & 6,997 & 0.2\% \\
\hline 
\end{tabular}
%\tablefoot{%Number of former satellites, $N_{\rm fs}$, and the corresponding 
%The numbers and percentages are separated into SF and Q galaxies.}
%\end{center}
\end{table}

It is important to note that for Q target galaxies of each model, a significant percentage were satellites in the past, which might be associated with the distance of these galaxies to other massive structures. \cite{Lacerna_2022} found that quenched low-mass central galaxies are typically much closer to massive halos than SF central galaxies of the same mass at $z=0$. 
However, SF target galaxies do not show significant percentages of former satellites even when these galaxies are located near massive halos as well. 
We measured that the median distance to the nearest massive system with M$_{\rm 200} \geq 10^{13}$ $h^{-1}$ M$_{\odot}$ at $z=0$ is 1.2 and 2.9 $h^{-1}$ Mpc for Q and SF target galaxies in MDPL2-SAG, whereas it is 2.3 and 3.3 $h^{-1}$ Mpc for Q and SF target galaxies in TNG300. Therefore, SF galaxies are farther than Q galaxies from massive halos, even for the target sample in both simulations.
We repeated the exact measurement for the former satellites of the target sample. We found that the median distance to the nearest massive system at $z=0$ is 1.2 and 1.3 $h^{-1}$ Mpc for Q and SF galaxies in the MDPL2-SAG model, and it is 1.6 and 2.3 $h^{-1}$ Mpc for Q and SF galaxies in TNG300. Here, the differences in the nearest distance to a massive halo are closer between Q and SF target galaxies that were satellites in the past, especially in the MDPL2-SAG model, but still quenched low-mass galaxies are typically closer to a massive structure than SF galaxies of the same stellar mass. 

Since the fraction of former satellites is low in the SF target samples of both catalogs, this suggests that most of the former satellites became quenched systems by environmental processes when they belonged to another halo. Therefore, they are ``polluting" the sample of central galaxies due to the quenching processes when their main progenitors were satellites.
Given that around half of the Q low-mass central galaxies in TNG300 were satellites in the past, we traced the evolution of ``pure'' low-mass central galaxies by removing these former satellites. With this, we aim to understand the role of these objects in the results of Sect. \ref{subsec_evolution}.
The results are as follows.

For TNG300, the main differences between the target sample before and after removing former satellites come from the host halo mass growth, v$_{\rm max}$, and the gas content for Q galaxies at $z \lesssim 1$. We show these parameters in Fig. \ref{fig:target_control_q_removing_fs_TNG}. Their median trends (red lines) show a more flattened evolution compared to the previous trends found in Fig. \ref{fig:all_params_target_TNG_and_SAG}, in which we had a drop during the last stages of their evolution. Therefore, the drop in halo and gas mass was produced by former satellites as a consequence of the environmental effects discussed above that affected them at that time. There is a drop in v$_{\rm max}$, but it occurs at a lower redshift and with a smaller amplitude than the case with all the Q target galaxies. A remarkable difference is that massive DM halos do not host Q target galaxies after removing former satellites.
If we compare how the trends look with respect to the control galaxies (gray lines) in Fig. \ref{fig:target_control_q_removing_fs_TNG}, we can still find slight differences at lower redshift, where Q target galaxies show lower median values than control ones. 

However, these differences are smaller than those in the top panels of Fig. \ref{fig:target_control_q_sf_TNG} before removing former satellites.
Once the former satellites are excluded, the stellar mass and sSFR parameters do not change whether the galaxy belongs to the target or the control sample. For SF galaxies, the median trends do not change significantly after removing former satellite galaxies. Moreover, the median trends for target and control SF galaxies are quite similar to those in Fig. \ref{fig:target_control_q_sf_TNG}, so we have decided not to include these figures. 
Therefore, at least for low-mass SF galaxies, the evolutionary histories of galaxies near massive systems are similar to those of galaxies in regions farther from massive groups or clusters.

For MDPL2-SAG, the median trends for Q and SF galaxies, either from the target or the control sample, do not show differences when former satellites are removed from the analysis (see Appendix  \ref{app:A}).
For SF galaxies, this behavior could be explained by the low fraction of former satellites (less than 2\%) on those samples. However, for Q galaxies the percentages are greater, reaching 17\% for Q target galaxies. We expected that by removing the former satellites in the target sample, the evolutionary trends of the gas and halo mass would be comparable to their control sample, as the results obtained from TNG300. However, the halo mass, v$_{\rm max}$, and gas mass still exhibit a drop at low redshift, leading to noticeable differences between the target and control samples. One explanation could be related to the fact that all former satellites found in the semi-analytic model were inhabiting less massive halos compared to those former satellites identified in TNG300. This behavior can be seen clearly in the lower-left panel of Fig. \ref{fig:all_params_target_TNG_and_SAG}. Moreover, this behavior is quite similar to the results obtained for the SF galaxies of TNG300, in which we do not have a large fraction of galaxies that inhabited massive halos to alter the median trends on their evolution. Then, its trends remain the same, which is what we have seen for all the galaxies in MDPL2-SAG.

In summary, the former satellite galaxies strongly influenced the evolution of the host halo mass, v$_{\rm max}$, and gas mass of the Q target galaxies in TNG300 shown in Fig. \ref{fig:all_params_target_TNG_and_SAG}. This is because most of their progenitors were hosted by massive DM halos, where environmental effects such as RPS, starvation, and TS could have affected those properties. The following section presents the analysis of their contribution to the signal of two-halo conformity. Based on the results of this section, we expect the Q target galaxies that were satellites in the past can explain a significant part of the conformity signal in TNG300, but it is less evident in the case of MDPL2-SAG.

%%%%%

%%%%%%%%%%%%%%%%%%%%%%%%%%%%%%%%%%%%%%%%%%%%%%%%%%%%%%%%%%%%%%%%%%%%
\section{Conformity signal} %\label{sub:conformity_signal}
\label{sec:conformity}
%%%%%%%%%%%%%%%%%%%%%%%%%%%%%%%%%%%%%%%%%%%%%%%%%%%%%%%%%%%%%%%%%%%%
In the literature, only a few studies have been carried out to analyze the influence of former satellites on the two-halo conformity signal, from which inconclusive results have been obtained on this matter, as we have previously mentioned. This section uses the samples of primary and secondary galaxies defined in Sect. \ref{sebsec_primary_secondary} to measure the conformity and, based on the findings in Sect. \ref{subsec:removing_fs}, analyze the contribution of former satellites to the signal in the local universe and at higher $z$. 

%%%
\subsection{Conformity signal in the local universe}
%%%

\begin{figure*}
\centering
\subfigure[\label{fig:conf_z0_TNG}]{\includegraphics[width=62mm]{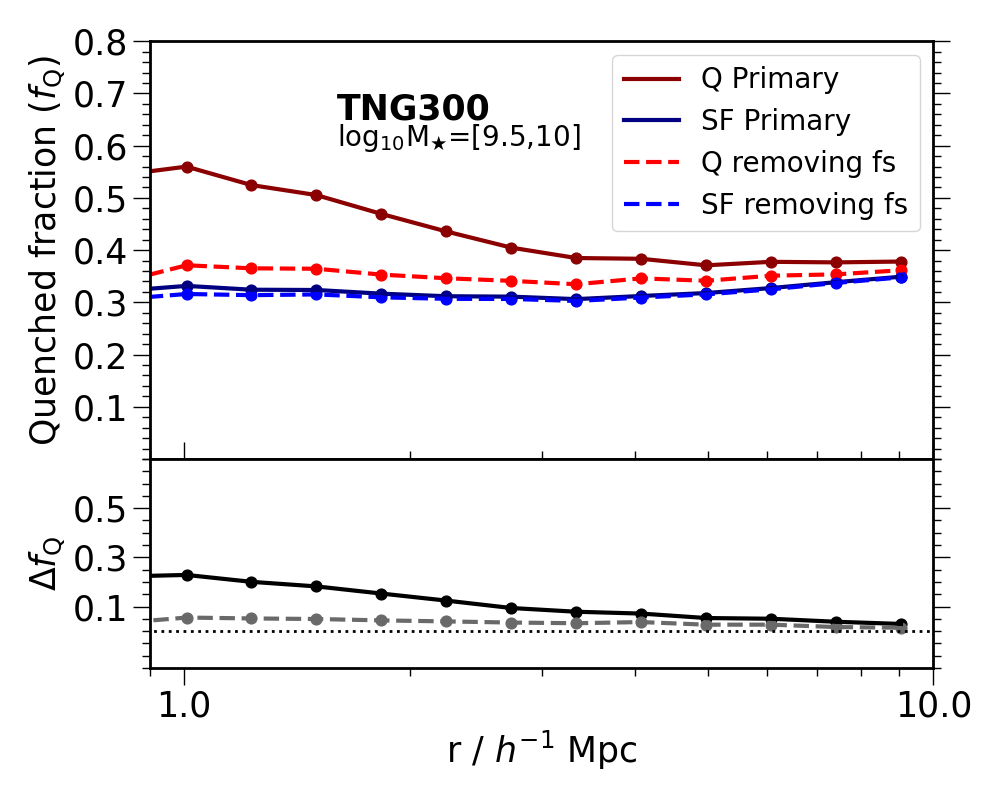}}\hspace{-1.85mm}
\subfigure[\label{fig:conf_z03_TNG}]{\includegraphics[width=62mm]{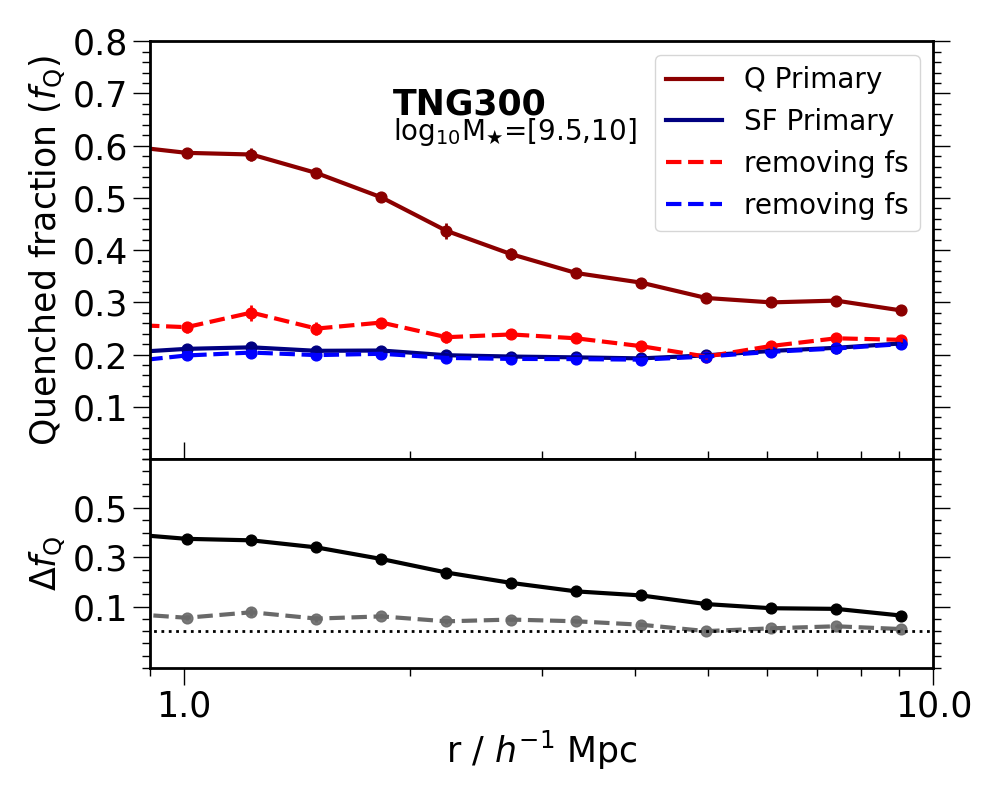}}\hspace{-1.85mm}
\subfigure[\label{fig:conf_z1_TNG}]{\includegraphics[width=62mm]{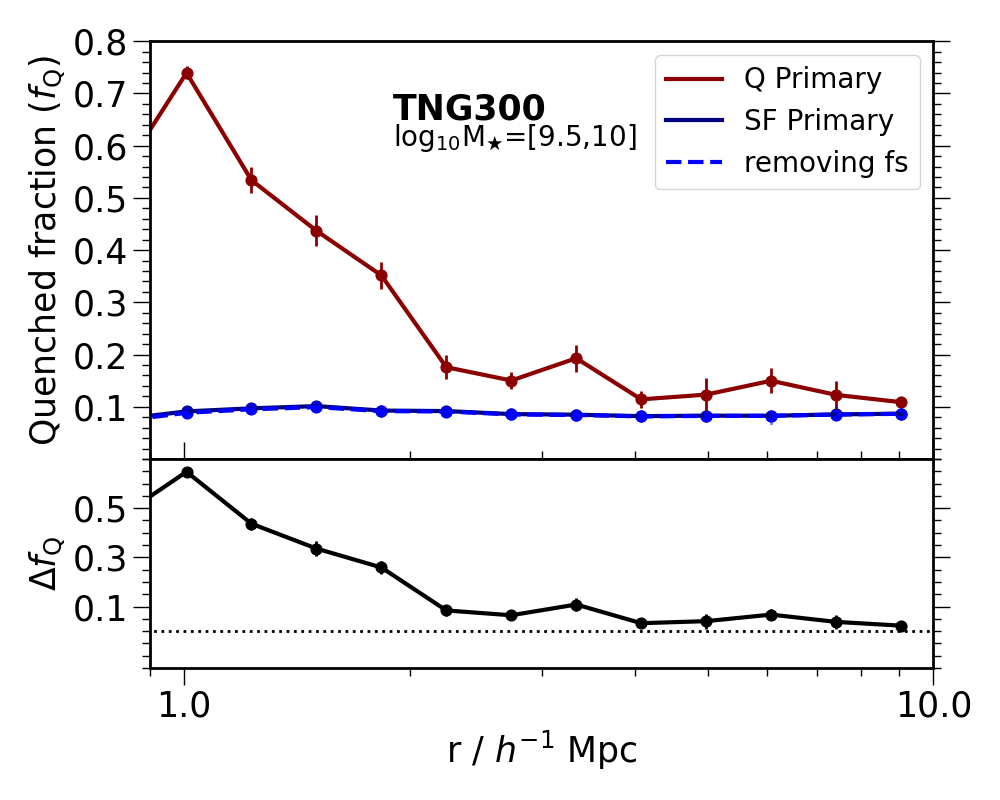}}
\caption{Conformity signal in TNG300 at $z$ = 0 (left), $z \sim 0.3$ (middle), and  $z \sim 1$ (right). The upper panels show the mean quenched fraction of secondary galaxies for both quenched (red) and star-forming (blue) primary galaxies before (solid lines) and after (dashed lines) removing former satellites. The bottom panels show how the conformity signal varies before (black solid line) and after (gray dashed line) removing former satellites, until 10 $h^{-1}$ Mpc distances from primary galaxies.
}
\label{fig:conf_signal_TNG}
\end{figure*}

Figure \ref{fig:conf_z0_TNG} shows the conformity signal at $z=0$ for TNG300.
The upper panel shows the mean quenched fraction ($f_{\rm Q}$) of all neighbor galaxies of the Q and SF primary samples (red and blue solid lines, respectively) until separations of 10 $h^{-1}$ Mpc. As discussed in Sect. \ref{section:former_satellites} and \ref{subsec:removing_fs}, our primary sample of central galaxies may be contaminated by former satellites (reported in Table \ref{tab:number_target_control_former_satellites_TNG_SAG}). Thus, we also estimate $f_{\rm Q}$ after removing former satellites from the primary sample (dashed lines).
Since the primary galaxies are target plus control galaxies, we have removed all former satellites regardless of whether they are target or control galaxies.
The lower panel shows the conformity signal, that is, the difference in the $f_{\rm Q}$ of all neighbor galaxies at a given distance from the Q and SF primary samples, before (black solid line) and after (gray dashed line) removing former satellites from the primary sample. We estimated the uncertainties in the mean of $f_{\rm Q}$ using the Jackknife method by dividing the simulation into 8 sub-boxes. For this, we calculated the standard deviation per unit bin distance by removing a single sub-box at a time. Therefore, the errors in the mean quenched fraction correspond to the standard deviation from the 8-sub boxes. Given the large number of secondary galaxies, the error bars are small enough to be imperceptible in some of these figures.

By construction, the solid lines in Fig. \ref{fig:conf_z0_TNG} correspond to the trends found in \cite{Lacerna_2022} for the same stellar mass range of primaries using TNG300, while the dashed lines correspond to the new results in this study. The latter is the ``clean sample'' of low-mass central galaxies without former satellites. In the lower panel, it can be seen that the signal is low at distances larger than 4 $h^{-1}$Mpc after removing former satellites from the primary sample. However, between 1 and 4 $h^{-1}$Mpc, there are signs of conformity, that is, the Q fraction of neighboring galaxies still depends on the sSFR of the primary galaxy. Thus, we could say that the signal at $z = 0$ is strongly influenced by former satellites in the past. However, these galaxies cannot fully explain it since the signal persists even when these galaxies are removed from the primary sample. \\

In our study, the former satellites in TNG300 contribute 75\% (1 - the ratio between gray and black lines in the lower panel of Fig. \ref{fig:conf_z0_TNG}) of the signal at 1 $h^{-1}$ Mpc and decrease from 70\% to 50\% between 2 and 4 $h^{-1}$ Mpc. This result can be compared to the findings obtained by \citet[][see their Fig. 6]{Wang_2023} using TNG300, in which the authors attributed all the two-halo conformity signal to former satellite galaxies. However, it is important to note that their study only considered central galaxies as neighbors (see Appendix \ref{app:B} for further discussion when using only central galaxies as neighbors), so the role of neighboring satellites was not included in the measured signal, which might explain the differences of 25--50\% in the contribution of former satellites compared with those authors. 
We have performed the test considering only central galaxies for the secondary sample at $z=0$, and the results (see Fig. \ref{fig:conformity_z0_only_central_as_neighbors}) showed their overall contribution is about 67\% of the conformity signal at scales between 1 and 2 $h^{-1}$ Mpc. 

\begin{figure}[!h]
  \centerline{\includegraphics[scale=0.32]{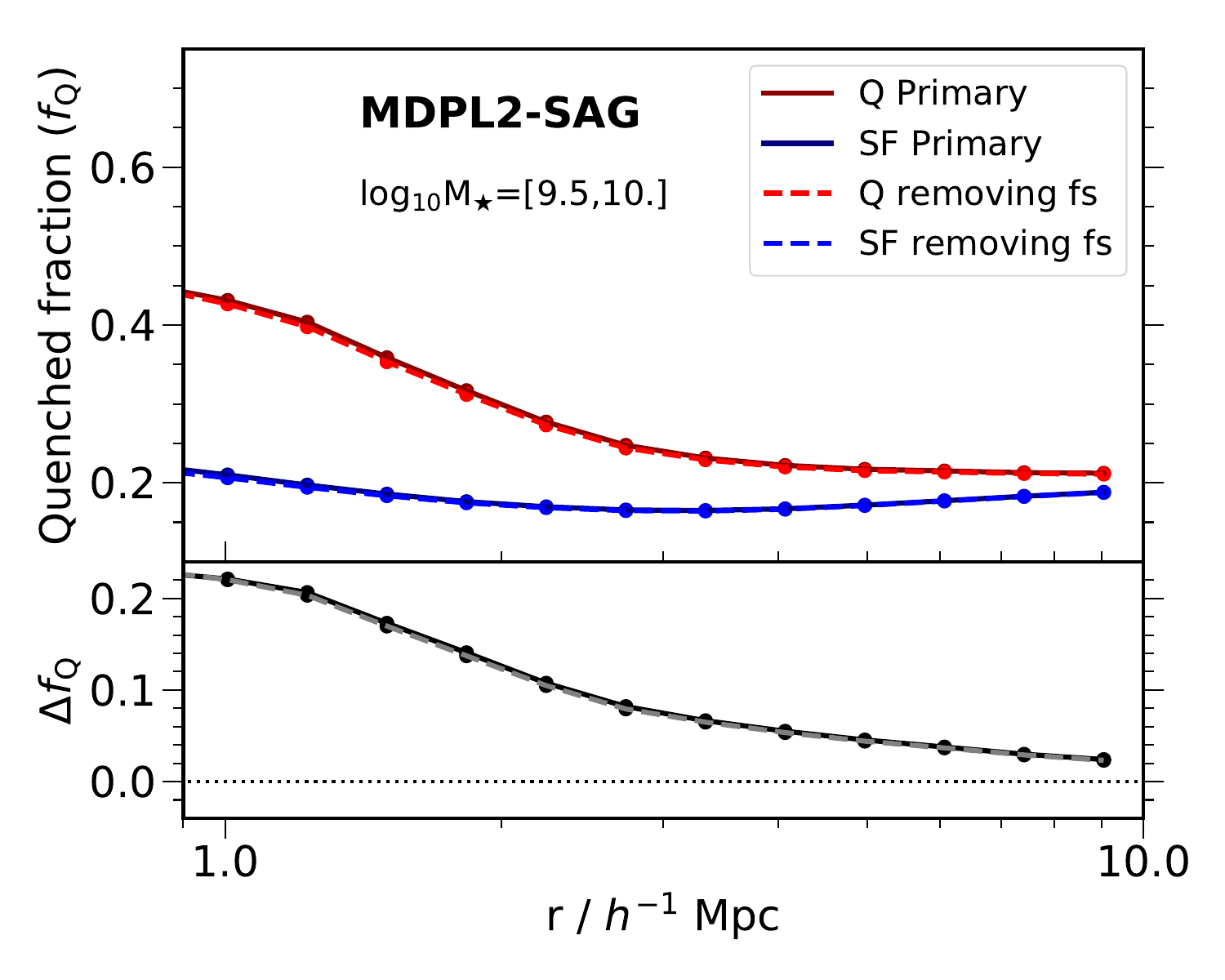}}
  \caption{Conformity signal in MDPL2-SAG at $z$ = 0. The lines are the same as those described in Fig \ref{fig:conf_signal_TNG}.\label{fig:conformity_signal_at_z0_SAG}}
\end{figure}

Figure \ref{fig:conformity_signal_at_z0_SAG} shows the mean quenched fractions of secondary galaxies around primary galaxies using the MDPL2-SAG model at $z=0$. 
The errors in $f_{\rm Q}$ are estimated using the diagonal of the covariance matrix after splitting every sample into
120 subsamples. Again, the error bars are very small, given the large number of secondary galaxies.
Although $f_{\rm Q}$ values for the fiducial cases (red and blue solid lines) differ from those in TNG300, they are qualitatively similar. Moreover, the conformity signal (black solid line in the lower panel) is quantitatively similar between both simulations, something already pointed out in \cite{Lacerna_2022}, but now using the same stellar mass range for the primaries. 
The conformity signal does not change after removing former satellite galaxies in the semi-analytic model (gray dashed line), though. This result is consistent with the unchanged evolution of the physical properties of low-mass central galaxies in MDPL2-SAG after removing the former satellites. 
Using a SAM (LGal-A21) model, \cite{Ayromlou_2023} claim that the contribution of former satellites to the conformity signal is less than 20\%, which is consistent with us using the MDPL2-SAG model in the sense that the contribution of former satellites in two different SAMs is low.

The two-halo conformity signal seems to depend strongly on the model and the identification of former satellites used. This dependency occurs even after carefully classifying this type of galaxies. % 
It is likely that the quantitative differences in the contribution of former satellites (``polluting" central galaxies) to the conformity signal occur because of the inherent nature of the simulations.  
The lower contribution in the SAMs compared to that observed in IllustrisTNG300 could be related to their specific technical properties, such as the reconstruction of merger trees or the identification of substructures. However, why this contamination is stronger in one model than another is beyond the scope of this paper.

%%%
\subsection{Conformity signal at higher redshifts}
\label{subsec:conformity_high_z}
%%%
We have seen that the two-halo conformity signal at $z=0$ is partly driven by former satellite galaxies in TNG300. These results evidence the importance of these galaxies in the measured signal. Although former satellites mostly explain the conformity signal, there is still some signal between 1 -- 4 $h^{-1}$Mpc that remains. Here, we analyze how the signal behaves at higher redshifts to visualize their impact along the evolution. We have selected two snapshots in our study, $z \sim 0.3$ and $z \sim 1$. This selection was made based on our previous results, in which we found interesting trends at $z \sim 0.3$ (halo mass) and $z \sim 1$ (gas mass), showing significant changes in the evolution of quenched, low-mass central galaxies in the outskirts of massive systems, whose behavior is associated with former satellites. At $z \sim 1$, in addition, the results can be directly compared with those from \cite{Ayromlou_2023}.

Figures \ref{fig:conf_z03_TNG} and \ref{fig:conf_z1_TNG} are similar to Fig. \ref{fig:conf_z0_TNG} but 
show the conformity signal at $z \sim 0.3$ and $z \sim 1$, respectively, for TNG300. From both figures, we can see that the signal (black solid line) increases significantly compared with that at $z=0$, which suggests that the growth of the conformity signal is a function of redshift, being in good agreement with \cite{Ayromlou_2023} for this simulation. At $z \sim 0.3$, the signal significantly decreases after removing former satellites. The gray dashed line shows some signal between 1 and 4 $h^{-1}$ Mpc, the same distance bin with a residual signal found at $z=0$. We measure that, at $z \sim 0.3$, former satellites contribute to the conformity signal up to 85\%. This result could tell us that some other environmental process could act on the conformity signal at these scales but with lower importance, at least in TNG300. On the other hand, if we analyze the signal at $z \sim 1$ and then remove the former satellites, the signal disappears because all quenched, low-mass central galaxies were identified as satellites at least once in the past ($z > 1$). This scenario was proposed by \cite{Wang_2023}, where they found that all the conformity signal at $z=0$ comes from these former satellite galaxies. However, as we have seen, the signal disappears completely at $z = 1$ but does not disappear completely at lower redshift. 

\begin{figure}
  \centerline{\includegraphics[width=8.5cm, height=6cm]{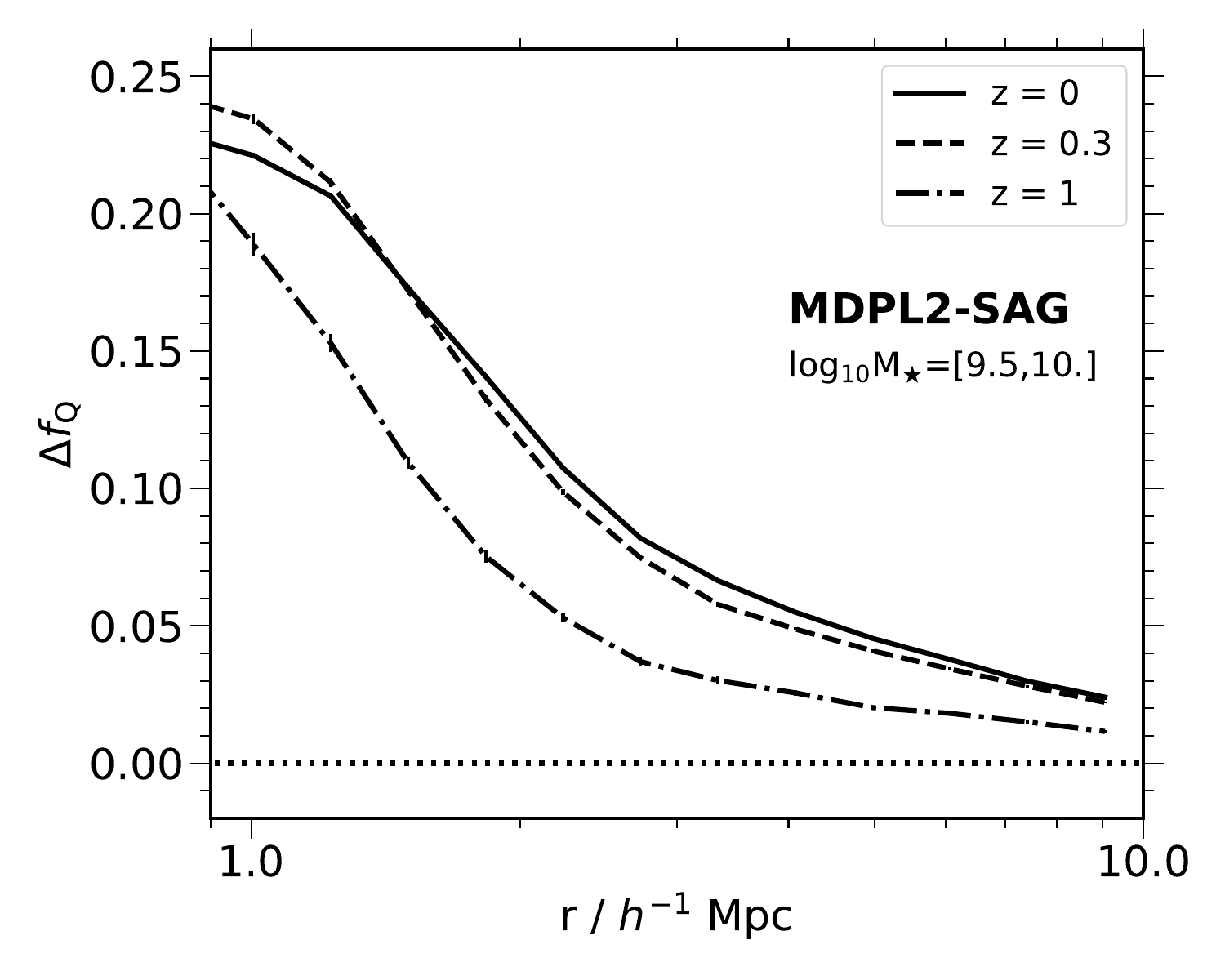}}
  \caption{Same as the lower panel of Fig. \ref{fig:conformity_signal_at_z0_SAG}. The conformity signal in MDPL2-SAG is shown at three different epochs: $z=0$ (solid line), $0.3$ (dashed line), and $1$ (dotted dashed line).\label{fig:conformity_comparison_SAG}}
\end{figure} %width=9.5cm, height=7cm

Figure \ref{fig:conformity_comparison_SAG} summarizes the fiducial cases for the conformity signal at three different redshifts in the semi-analytic model. We have decided to compare in a single figure the fiducial conformity signals in MDPL2-SAG since former satellites in this model do not contribute to the signal. At $z=0.3$ (dashed line), there is a slight enhancement of the signal at scales smaller than 2 $h^{-1}$ Mpc with respect to the case at $z=0$ (solid line). However, at distances larger than 2 $h^{-1}$ Mpc from primary galaxies, both lines are inverted, showing more correlation at present. If we analyze the signal at $z=1$ (dotted dashed line), we can see the signal decreases over the whole range of distances. This result contradicts the findings in TNG300 for the fiducial case, in which we found the signal grows as a function of redshift. 
Still, there is a better agreement when the former satellites are removed in the hydrodynamic simulation. From the literature, we can compare our results with \cite{Ayromlou_2023}. Using an updated version of the L-Galaxies semi-analytic model, they found the signal decreases from $z=0$ to $z=2$, showing an enhancement at $z=1$ compared with $z=0$. In our case, the smaller amplitude of the conformity signal in the MDPL2-SAG model at $z=1$ than that in the present is comparable with their findings using the EAGLE simulation. As discussed by those authors, the time evolution of the signal is complex, and the different results might come from the impact of the physical properties involved in each model. We also point out the contribution of former satellites, given by the construction and characteristics of each simulation, in their respective evolution of the conformity signal.

%%%%%%%%%%%%%%%%%%%%%%%%%%%%%%%%%%%%%%%%%%%%%%
\section{Discussion and conclusions}
\label{sec:discusion_conclusions}
%%%%%%%%%%%%%%%%%%%%%%%%%%%%%%%%%%%%%%%%%%%%%%

We know from previous studies (e.g., \citealt{Lacerna_2022}, \citealt{Ayromlou_2023}) that low-mass central galaxies (M$_{\star}$ = 
[10$^{9.5}$,10$^{10}$] $h^{-1}$ M$_{\odot}$) in the vicinity of galaxy groups and clusters can produce a strong two-halo conformity signal, that is, the correlation in sSFR between the central galaxy and the neighboring galaxies (either central or satellite galaxies). We   analyzed how the physical properties of these low-mass central galaxies (called target galaxies) evolve from $z=2$ to the present using the hydrodynamical simulation IllustrisTNG300 and the semi-analytic galaxy catalog MDPL2-SAG to find hints regarding the origin of this signal. To this end, we   separated the galaxies into quenched (Q) and star-forming (SF) galaxies at $z=0$. We   also considered central galaxies of the same stellar mass located farther from the center of massive systems (called control galaxies). In addition, we   identified a subsample of former satellite galaxies, that is, central galaxies at present, but that were satellites in the past, that pollute our central galaxy samples because they may have suffered other environmental processes when they were satellites in other dark matter halos. We found that former satellites in IllustrisTNG300 can influence the median trends in the overall evolution of low-mass central galaxies and can contribute to the conformity signal.
Our main results are as follows:
\begin{itemize}
    \item The evolution of halo mass, v$_{\rm max}$, and gas mass parameters for low-mass quenched central galaxies in the vicinity of massive groups and clusters reveals a significant drop in their median values at $z \lesssim 1$, attributed to the presence of the subsample of former satellites in TNG300, which are susceptible to environmental effects. We also find a decline in halo mass, v$_{\rm max}$, and gas mass in quenched target galaxies of MDPL2-SAG; however, the drop here is not associated with former satellites.
    The drop in halo mass for Q target galaxies is not observed in the control sample of quenched galaxies in the two simulations.
    For SF galaxies, instead, no major differences are found between target and control galaxies during their evolution.
    
    \item After removing the subsample of former satellites on TNG300, the median trends of halo mass, gas mass, and v$_{\rm max}$ for Q target galaxies exhibit similar trends compared to Q control galaxies (Fig. \ref{fig:target_control_q_removing_fs_TNG}). However, small differences arise at lower redshifts, showing slightly lower median values in these parameters for target galaxies, with a flatter behavior from $z < 0.5$ for halo mass and gas mass. In contrast, galaxies from MDPL2-SAG show no evident variations in their evolutionary histories once the subsample of former satellite galaxies was removed. Therefore, the differences found between Q target and Q control galaxies remain for this catalog.
    
    \item The percentages of former satellites for both TNG300 and MDPL2-SAG are higher for Q galaxies, reaching 45\% and 17\% for Q target galaxies and less than 7\% for SF target galaxies. 
    Most former satellites in Q target galaxies appear to be hosted by massive dark matter halos in TNG300, with mean halo masses around $10^{13.5}$ $h^{-1}$ M$_{\odot}$. However, the scenario in MDPL2-SAG is quite different, showing that all former satellites are hosted by less massive halos, with mean halo masses around $10^{11.4}$ $h^{-1}$ M$_{\odot}$.
    
    \item Former satellites play an important role in the two-halo conformity signal in TNG300. At higher redshifts ($z \sim 1$), these galaxies can explain the whole signal, while at lower redshifts ($z$ $\lesssim$ 0.3), they contribute up to 75 -- 85\%. 
    Therefore, there is a remaining signal, around 15 -- 25\%, in the clean sample of low-mass central galaxies that could be associated with other processes at $z$ $\lesssim$ 0.3 in this simulation. Based on our results, the contribution of former satellites in the two-halo conformity is lower for SAMs than the hydrodynamical model IllustrisTNG300. Using MDPL2-SAG, we found a negligible contribution of this type of galaxy on the signal.

    \item{The time evolution of the conformity signal at megaparsec scales is apparently contradictory in both simulations. The signal decreases from $z=0$ to $z=1$ in MDPL2-SAG, while it increases in TNG300. However, after removing former satellites in the latter, the signal practically does not change at $z \leq 0.3$, and it disappears at $z=1$.}

\end{itemize}

The TNG300 and MDPL2-SAG synthetic galaxy catalogs are very different because the physical processes involved may act in different ways on the galaxies, or the efficiency of these processes is not the same. Thus, we expected to find quantitative differences in the evolution of low-mass galaxy properties, such as stellar mass, gas mass, and sSFR. Nonetheless, the results presented in this study show that the evolution of these properties is qualitatively similar in the two simulations (see Figs. \ref{fig:all_params_target_TNG_and_SAG}--\ref{fig:target_control_q_sf_SAG}). Differences in the numerical resolution of each simulation lead to different fractions of former satellites in each model, where this fraction increases as the resolution is higher.
The inherent differences between TNG300 and MDPL2-SAG are manifested in the different mean quenched fractions of neighbor galaxies around the low-mass central objects at $z = 0$. However, the two-halo conformity signal, that is, the difference in the quenched fraction of neighbor galaxies at a given distance from the Q and SF low-mass centrals, is very similar in the two simulations 
(black solid lines in Figs. \ref{fig:conf_z0_TNG} and \ref{fig:conformity_signal_at_z0_SAG}).
Therefore, the results show that a similar conformity signal is obtained in the two galaxy catalogs by different physical processes. As \cite{Lacerna_2022} and \cite{Ayromlou_2023} demonstrate, most of this signal comes from central galaxies near massive galaxy groups and clusters, that is, the target galaxies. The quenched target galaxies produce an excess of correlation between them and quenched neighboring galaxies. The question arises of why those low-mass target galaxies are quenched. 
After checking the evolution histories of their physical properties, including those related to their host DM halo, and comparing them with the low-mass central galaxies of control, we can conclude that there is no single explanation for the quenching of target galaxies.

We find  in the case of TNG300 that 75--85\% of the two-halo conformity signal comes from target galaxies that were hosted by massive dark matter halos in the past. According to \cite{Ayromlou_2023}, former satellite galaxies can lose gas and dark matter due to stripping inside and in the outskirts of their former host halos. We add that this is more likely to occur for low-mass central galaxies that were satellites in massive halos \citep[see also][]{Wetzel_2014}, which naturally explains the quenching in their star formation activity. Our results for the two-halo conformity in TNG300 are qualitatively consistent with the results found in \cite{Wang_2023} using the same model. Although they found that the signal could be explained entirely by central galaxies that were satellites in the past, they used only central galaxies as secondary galaxies.  
We contrast our results with their study (Appendix \ref{app:B}), and we found that part of the conformity signal comes from the secondary sample of satellite-type galaxies (not only from central neighbors) around Q primary galaxies.
In addition, we tested the importance of the former satellite definition, which should be considered carefully. 
It can lead to different results, due to cases of misclassification of former satellite galaxies by the halo finder.

We find an absence of low-mass central galaxies accreted by massive halos in the past in MDPL2-SAG. The former satellites do not contribute to the two-halo conformity measured in this catalog.
Compared with the study performed by \cite{Ayromlou_2023} using the SAM LGal-A21, both SAMs show that the former satellite galaxies contribute less than 20\% to the conformity signal. Since we found a negligible contribution of former satellites in MDPL2-SAG, we attribute the quantitative differences between these SAMs to the prescriptions applied in each model and the identification of former satellites in the catalogs.

Therefore, the most likely explanation for the excess of correlation in sSFR between quenched target galaxies and their quenched neighbor galaxies in TNG300 is not applied to the targets in MDPL2-SAG. 
The remarkable drop in M$_{\rm 200}$ since $z \sim 0.46$ for Q targets, followed by the drops in M$_{\rm gas}$ and sSFR since $z \sim 0.43$ and $0.28$, respectively (Fig. \ref{fig:target_control_q_sf_SAG}), suggest that the tidal stripping of halo mass by massive neighbor halos \citep{Hahn_2009} can be responsible for the quenching of the target galaxies in MDPL2-SAG, which in turn may explain the two-halo conformity in this catalog.
Since the host halos of central target galaxies have not been part of a larger system before, the severe mass stripping could also be due to tidal forces exerted by the `cosmic web' \citep{Borzyszkowski_2017, Montero-Dorta_2024}. 
For example, tidal forces in galaxy clusters are quite efficient at disrupting smaller systems, such as groups, by removing their satellites (e.g., \citealt{Choque-Challapa_2019}, \citealt{Haggar_2023}), so we might expect something similar to happen in dense large-scale structures such as thick filaments.
An in-depth study of the role of filaments in the mass stripping of low-mass halos in the vicinity of groups and clusters will be addressed elsewhere.

\begin{acknowledgements}
We thank the referee for their comments that helped to improve this work.
We would like to thank Nelson Padilla, Tom\'as Hough, Vladimir Avila-Reese, Sergio Contreras, and Yetli Rosas-Guevara for useful comments and discussions. We also thank Yamila Yaryura and Cristian Vega for their support in organizing and providing the MDPL2-SAG data used in this work. DP acknowledges the support through ANID-Subdirección de Capital Humano/Doctorado
Nacional/2024/21241817. MCA acknowledges support from ANID BASAL project FB210003. ADMD thanks Fondecyt for financial support through the Fondecyt Regular 2021 grant 1210612. SAC acknowledge funding from {\it Consejo Nacional de Investigaciones Científicas y Tecnológicas} (CONICET, PIP-2876), and {\it Universidad Nacional de La Plata} (G11-183), Argentina. FR thanks the support by Agencia Nacional de Promoción Científica y Tecnológica, the Consejo Nacional de Investigaciones Científicas y Técnicas (CONICET, Argentina) and the Secretaría de Ciencia y Tecnología de la Universidad Nacional de Córdoba (SeCyT -UNC, Argentina). FR would like to acknowledge support from the ICTP through the Junior Associates Programme 2023-2028. DP acknowledges financial support from ANID through FONDECYT Postdoctorado Project 3230379. DP, gratefully acknowledges support by the ANID BASAL project FB210003. NCC acknowledges a support grant from the Joint Committee ESO-Government of Chile (ORP 028/2020). 
\end{acknowledgements}



%\clearpage
 
\bibliographystyle{aa}
\bibliography{aa50976-24} % if your bibtex file is called example.bib

%%%%%%%%%%%%%%%%%%%%%%%%%%%%%%%%%%%%%%%%%%%%%%%%%%

%%%%%%%%%%%%%%%%% APPENDICES %%%%%%%%%%%%%%%%%%%%%

%%%%%%%%%%%%%%%%%%%%%%%%%%%%%%%%%%%%%%%%%%%%%%%%%%

\begin{appendix}

\section{Evolution of central galaxies after removing former satellites in MDPL2-SAG}
\label{app:A}

\begin{figure}[!h]
\centering
\subfigure{\includegraphics[scale=0.195]{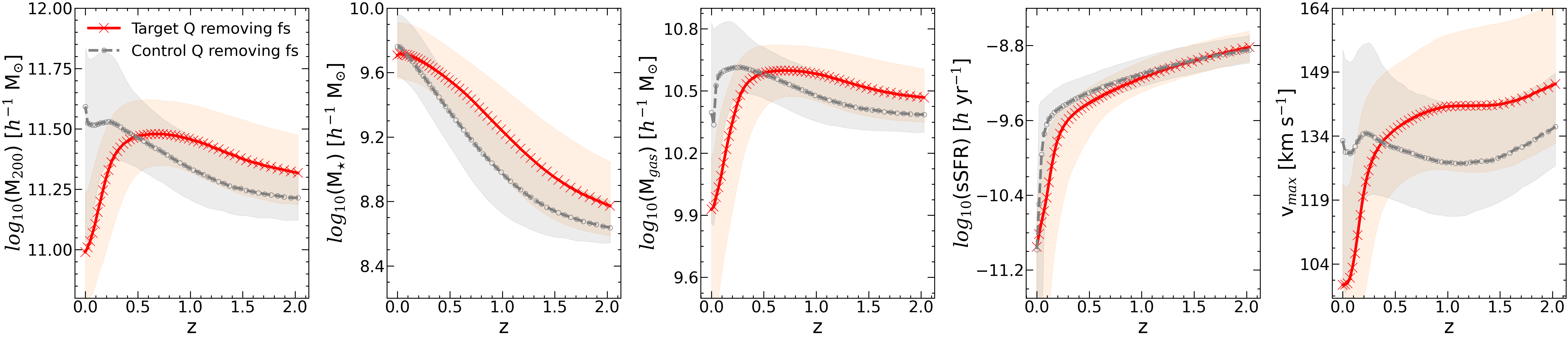}} %\par\smallskip
\\[-1.7ex]
\subfigure{\includegraphics[scale=0.195]{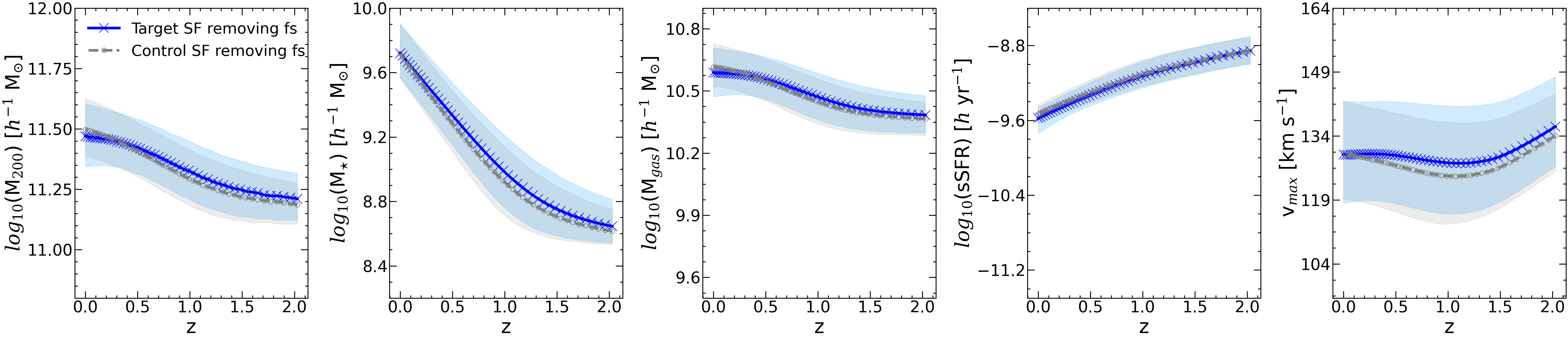}}
\onecolumn
\caption{Same parameters as Fig. \ref{fig:target_control_q_sf_SAG}. The upper panels show the median trends for Q galaxies (target in red, control in gray), and the lower panels show the SF galaxies for the target (blue) and control (gray) samples in MDPL2-SAG after removing former satellites.}
\twocolumn
\label{fig:target_control_q_sf_SAG_removing_fs}
\end{figure}

For completeness, Fig. \ref{fig:target_control_q_sf_SAG_removing_fs} shows the evolution in the properties of low-mass central galaxies in the MDPL2-SAG catalog after removing former satellites. The trends are quite similar to those shown in Fig. \ref{fig:target_control_q_sf_SAG}.

% SAG after removing FS

\FloatBarrier

\clearpage

%%%%%%%%%%%%%

\section{Two-halo conformity considering only central galaxies as the secondary sample}
\label{app:B}

\begin{figure}[h!]
%\centerline{\includegraphics[scale=0.37]{results/conf_z0_only_centrals_as_neighbors.png}}
\centerline{\includegraphics[scale=0.32]%[scale=0.37]
{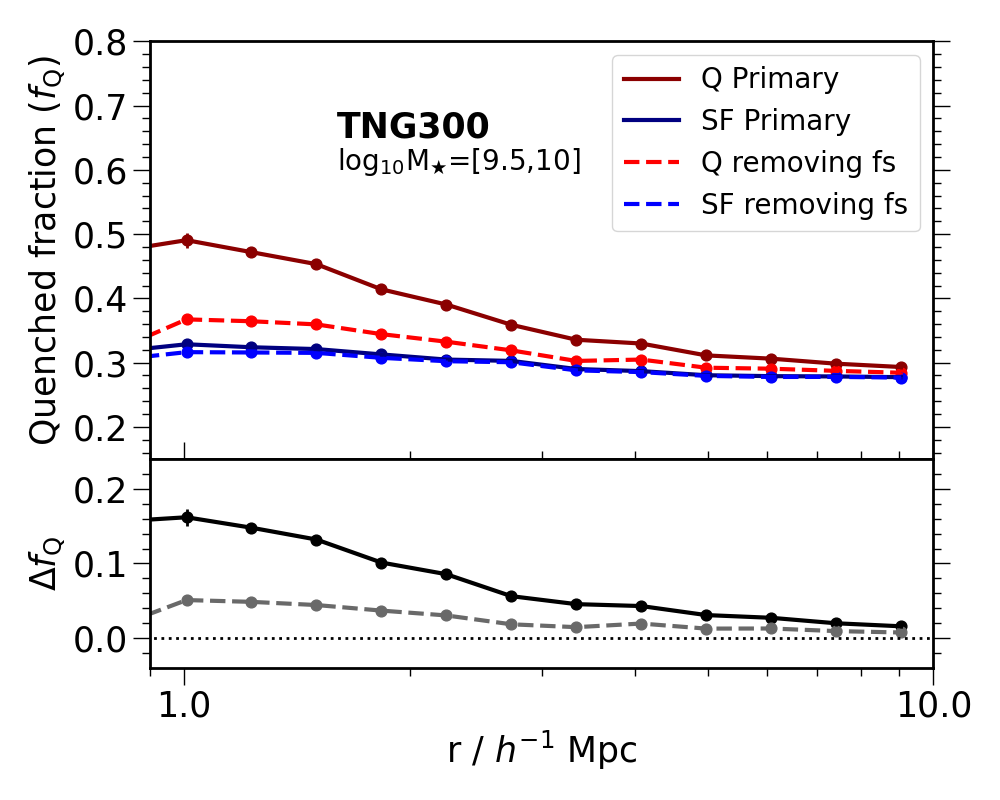}}
\caption{Same as Fig. \ref{fig:conf_z0_TNG}, but considering only centrals as neighbor galaxies, as in \cite{Wang_2023}.
\label{fig:conformity_z0_only_central_as_neighbors}}
\end{figure}

As mentioned in the text, \cite{Wang_2023} presented results where the former satellites can explain all the two-halo conformity signal in TNG300 at $z \sim 0$. A difference in that work compared to others is that they used only central galaxies in the sample of secondary galaxies. Figure \ref{fig:conformity_z0_only_central_as_neighbors} shows the results using that methodology, that is, considering only central galaxies as the neighbor sample around primary galaxies.
Compared to Fig. \ref{fig:conf_z0_TNG}, the $f_{\rm Q}$ for the Q primaries in the fiducial case (red solid line) is lower, while $f_{\rm Q}$ for the SF primaries (blue solid line) is similar, which produces a lower two-halo conformity signal (black solid line). The $f_{\rm Q}$ values shown in solid lines in Fig. \ref{fig:conformity_z0_only_central_as_neighbors} are consistent with those reported by \cite{Wang_2023} in their Fig. 6 for the same stellar mass range.
The conformity signal also decreases when the former satellites are not considered in the primary sample (gray dashed line), but it is not completely absent between 1 and 2 $h^{-1}$ Mpc in contrast to the results of those authors. We find that the former satellites contribute 67\% of the two-halo conformity signal at these scales. 

The difference in the contribution of former satellites to the conformity signal between this study and  \cite{Wang_2023} likely stems from the methodology used to select them.
In their case, the main progenitors of these central galaxies were once satellites of other halos and the satellite state lasts at least two successive snapshots. We checked there were cases in which a central galaxy at $z=0$ was clearly a satellite in the past just in a single snapshot, so we preferred not to use the condition of the satellite status in two successive snapshots. In our case, we considered a distance condition in which, at a given snapshot, a satellite candidate at $d \leq$ 2 R$_{200}$ from the host halo of the central galaxy is considered a former satellite. Otherwise, it is considered central because in TNG300 a galaxy can be associated with two different halos simultaneously in the same snapshot, either by late stages of a merger or ejected cores (see, e.g., \citealt{Poole_2017}), with different distances from the host halos, one at the center and the other at $d$ > 0. 
In both cases, the subhalo finder can flag it as central.  
Then, in the following snapshot(s), the galaxy can be flagged as a satellite at $d$ > 0.
We noticed that in a given snapshot where the satellite candidate is at $d >$ 2 R$_{200}$, and after checking its evolution in the previous and following snapshots, it is likely an artifact of the subhalo finder that flagged it as a satellite since it never crossed the virial radius of another halo. Therefore, in those cases, we flagged the galaxy as central. \cite{Wang_2023}, on the other hand, might have considered these galaxies as former satellites. Since these central galaxies are in the vicinity of massive structures, they might contribute significantly to the conformity signal, as shown in \cite{Lacerna_2022}. This could explain why these authors concluded that the entire signal originates from backsplash objects.
We consider our criteria for identifying former satellites to be reliable, and this is supported by checking that the evolution of target galaxies in TNG300 is more similar to that of control galaxies when the former satellites are removed (Sect. \ref{subsec:removing_fs}).

\end{appendix}

\end{document}